\documentclass[aps,pra, preprint,showpacs]{revtex4-1}
\usepackage{graphicx}
\usepackage{amsmath}
\usepackage{amsfonts} 
\usepackage{bm}
\usepackage{color}
\usepackage{subfigure}
\usepackage{txfonts}
\usepackage{setspace}
\usepackage{ulem}

\newcommand{\be}{\begin{eqnarray}}
\newcommand{\ee}{\end{eqnarray}}
\begin{document}

\title{Amplitude~/~Higgs Modes in Condensed Matter Physics}
\author{David Pekker$^1$ and C. M. Varma$^2$}
\affiliation{$^1$Department of Physics and Astronomy, University of Pittsburgh, Pittsburgh, PA 15217}
\affiliation{$^2$Department of Physics and Astronomy, University of California, Riverside, CA 92521}
\date{\today}

\begin{abstract}
The order parameter and its variations in space and time in many different states in condensed matter physics at low temperatures are described by the complex function $\Psi({\bf r}, t)$.  These states include superfluids, superconductors, and a subclass of antiferromagnets and charge-density waves. The collective fluctuations in the ordered state may then be categorized as oscillations of phase and amplitude of $\Psi({\bf r}, t)$. The phase oscillations are the {\it Goldstone} modes of the broken continuous symmetry. The amplitude modes, even at long wavelengths, are well defined and decoupled from the phase oscillations only near particle-hole symmetry, where the equations of motion have an effective Lorentz symmetry as in particle physics, and if there are no significant avenues for decay into other excitations. They bear close correspondence with the so-called {\it Higgs} modes in particle physics, whose prediction and discovery is very important for the standard model of particle physics. In this review, we discuss the theory and the possible observation of the amplitude or Higgs modes in condensed matter physics -- in superconductors, cold-atoms in periodic lattices, and in uniaxial antiferromagnets. We discuss the necessity for at least approximate particle-hole symmetry as well as the special conditions required to couple to such modes because, being scalars, they do not couple linearly to the usual condensed matter probes.
\end{abstract}
\maketitle

\section{Introduction}
Ever since the phenomenological theory of Ginzburg and Landau in 1950~\cite{Ginzburg1950}, it has been known that the long-range order of superfluids and superconductors and its slow variations in space and time must be described by a complex function describing an amplitude and a phase, i.e., a $\text{U}(1)$ matter field:
\be
\label{matter}
\Psi ({\bf r},t) = |\Psi({\bf r},t)| e^{i\phi ({\bf r},t)}.
\ee
This is in accord with the hydrodynamics introduced by Landau~\cite{Landau1941} for liquid $^4$He and the microscopic weak-coupling theory for Bosons by Bogoliubov~\cite{Bogoliubov1947, *Bogoliubov1947b}. The microscopic theory of superconductivity was invented by Bardeen, Cooper and Schrieffer~\cite{Bardeen1957} (BCS) in 1957 and almost immediately the phenomenological Ginzburg-Landau Lagrangian with (\ref{matter}) as the order parameter field was derived by Gorkov~\cite{Gorkov1959} in the static long wavelength limit of the theory. 

We are concerned in this review with a number of different systems with different microscopic physics but with long-range order and its long wavelength fluctuations describable by a field $\Psi(r,t)$. We therefore first write down the action-density for the field $\Psi(r,t)$ based on general symmetry principles and in subsequent sections show how the microscopic physics in different systems leads to it. This allows us to emphasize the generality of the ideas most simply and economically.

The Ginzburg-Landau action-density in the static and long wavelength limit is
 \be
\label{action}
S_{static} = - r \, \Psi^* \Psi + \frac{U}{2} \, (\Psi^*\Psi)^2 +  \xi^{-2}\, (\nabla \Psi^*)(\nabla \Psi) 
 \ee
For a charged matter field $\Psi({\bf r},t)$, as in superconductivity, one must include the action of the electromagnetic field and its interaction with $\Psi({\bf r},t)$ in a gauge-invariant way by changing $\nabla \to (\nabla -i (e/c) {\bf A})$, where ${\bf A}({\bf r}, t)$ is the vector potential. 
The effect of such a term is briefly mentioned in Sec.~\ref{sec:phenomenology}.
\begin{figure}
\includegraphics[width=0.6\textwidth]{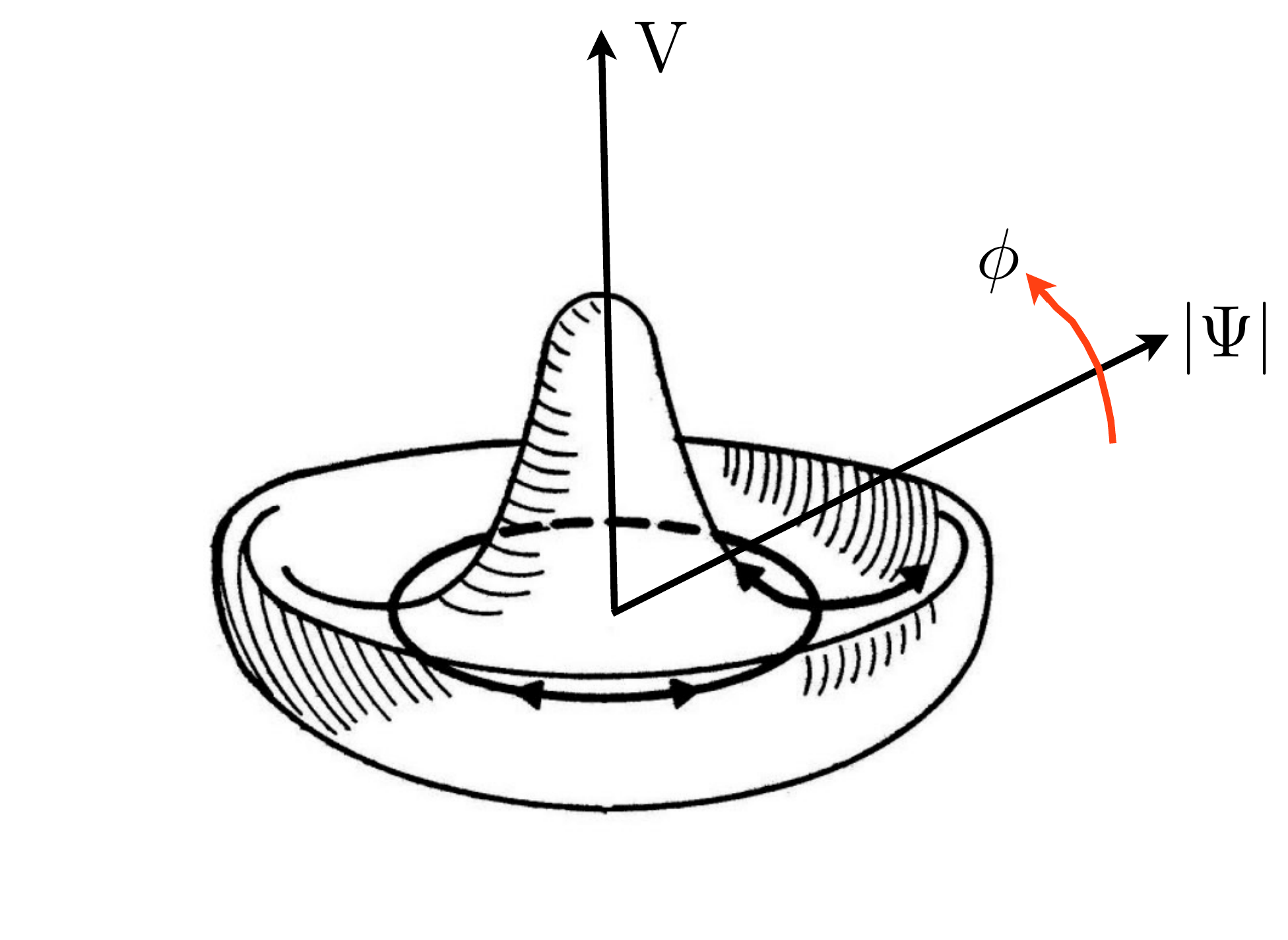}
\caption{The ``Mexican Hat Potential" $V(|\Psi|, \phi)$ for the condensate field $\Psi$ in the long wave-length limit as a function of the amplitude $|\Psi|$ and the phase $\phi$. Adapted from Reference~\cite{Varma2002}}
\label{Fig:mexhat}
\end{figure}

We need consider only the first two terms of Equation \eqref{action} in equilibrium. Then for $r >0$, the potential energy is represented pictorially as in Fig. (\ref{Fig:mexhat}). The static equilibrium is at 
\be
\label{psieq}
|\Psi_0| = \langle \Psi \rangle =  \sqrt{r/U}.
\ee
The equilibrium energy does not depend on the phase $\phi$ and we may pick it to be 0. 

To calculate the time-dependent fluctuations about this state one must supplement Equation~\eqref{action} with dynamical terms. Throughout this paper, we restrict ourselves to phenomena at low temperatures compared with the transition temperature, where loss of energy in the degrees of freedom in $\Psi({\bf r},t)$ compared with other degrees of freedom such as fermions, phonons, etc., is unimportant. The dynamical terms must then be time-reversal invariant. The first two time-reversal and  gauge-invariant terms allowed in the action-density are
\be
\label{s-dynamic}
S_{dynamic} = i K_1\Psi^*({\bf r},t) \frac{\partial}{\partial t}\Psi({\bf r},t) - K_2 \Big(\frac{\partial}{\partial t}\Psi^*({\bf r},t)\Big)\Big(\frac{\partial}{\partial t}\Psi({\bf r},t)\Big)
\ee
The distinction between the two terms is important in the subsequent discussion. On deriving the equation of motion from $S_{static} + S_{dynamic}$ (as detailed in the next section), one observes that retaining the $K_1$ term alone gives time dependence, as in the Schr\"{o}dinger equation. This is what occurs, appropriately enough, for the dynamical equations for superfluid helium given by Gross~\cite{Gross1961} and Pitaevskii~\cite{Pitaevskii1961}, which are not Lorentz invariant. In the present context, it is more important to emphasize the particle-hole symmetry aspect of Lorentz invariance because it requires that the equation of motion be symmetric under conjugation. If the $K_1$ term is absent, e.g., owing to the presence of particle-hole symmetry, only the $K_2$ remains and a Lorentz-invariant equation of motion is obtained.

The equations of motion with both $K_1$ and $K_2$ terms included are derived in the next section, but we can say some useful things by looking at Fig. (\ref{Fig:mexhat}) alone and making some general considerations. The invariance of the energy to the phase implies that there must exist a zero energy or massless mode of fluctuation of density current in the long wavelength limit, i.e., along the azimuthal direction at the minimum of the potential in Fig. (\ref{Fig:mexhat}). This is the physical content of the Goldstone theorem~\cite{Goldstone1961,*Goldstone1962}. These are the sound modes derived for superfluid $^4$He and for the weakly interacting Bose gas. Due to the long-range nature of the Coulomb interaction, density or longitudinal fluctuations of the charge density in superconductors at long wavelengths occur not at zero energy but close to the plasmon energy as in the normal metallic state~\cite{Anderson1958}. These results are true whether or not the dynamics is Lorentz invariant or particle-hole symmetric. 

What about the orthogonal degree of freedom exhibited in Fig. (\ref{Fig:mexhat}), i.e., the oscillations of the amplitude of the order parameter $|\Psi({\bf r},t)|$ around its equilibrium value (\ref{psieq})? This is the Higgs or amplitude mode of the model. Its existence was first mentioned in a paper by Higgs~\cite{Higgs1964} for the same model but with Lorentz invariance, i.e., with a second-order derivative in time only, because he was interested in applications to particle physics. Its occurrence in superconductivity  was missed until 1981~\cite{Littlewood1981, *Littlewood1982}, because collective fluctuations of amplitude were studied primarily only near $T_c$, where the dominant time-dependent term describes the relaxation dynamics of the order parameter associated with coupling to the fermion bath. In this case, there is no distinct Higgs mode~\cite{Varma2002}. In superfluid $^4$He, $K_1 \ne 0$ and so no Higgs mode can be found either~\cite{Varma2002}. In superconductors at low temperatures, as is explained below, $K_1 \approx 0$ and $K_2 \ne 0$; the theory is Lorentz invariant to a very good approximation. So an amplitude mode or Higgs boson can exist provided it does not have avenues for rapid decay into quasiparticle-quasihole pairs.  Particle-hole symmetry occurs for interacting Bosons in a periodic lattice~\cite{Huber2007} along a line in their phase diagram and such a system was experimentally realized by cold atom techniques and the observations interpreted in terms of the Higgs or amplitude modes. Experiments in planar antiferromagnets have also been interpreted similarly. So have the experiments in the recently achieved exciton condensates~\cite{Littlewood2007}.

We start with very briefly discussing the amusing history of the relationship of developments in superconductivity and in the standard model of electro-weak unification in particle physics. In the following section, we discuss the phenomenology of the Higgs mode that appears in different physical systems under appropriate conditions. In subsequent sections, the microscopic theory and experiments in superconductors, cold bosonic atoms, and quite briefly in antiferromagnets are discussed. We place some emphasis on the fact that Higgs modes are quite hard to discover. They are directly discoverable in some cases -- as, for example, in superconductors -- only because the existence of a condensate allows some conservation law to be circumvented. We also mention a new way of discovering Higgs through Josephson-coupling effects.

\subsection{Brief History}

The story of the amplitude or Higgs mode, and the phase or Goldstone or Bogolubov mode in a gauge theory of chargeless matter, which moves to the plasma frequency in a gauge theory of charged matter, is among important matters, as is the story of cultural differences among condensed matter physicists and elementary particle physicists. The prolific and wide variety of experiments available in condensed matter physics compared with that in particle physics make certain things obvious that are less so in the experiment starved and therefore necessarily more abstract field of particle physics. Among condensed matter physicists,  there appears to be some confusion about why the Higgs discovered in particle physics is so important as well as about the meaning and difference of the terms, Higgs field, Higgs condensate, Higgs bosons, Bogolubov modes, Anderson-Nambu-Goldstone bosons.  We present the development of the idea of the Higgs or amplitude mode in condensed matter physics and along the way hopefully also clear up some of the confusion. We start by first giving a glimpse into the interesting history of this field.

BCS had chosen a gauge to do the calculations in which the transverse response of the superconductor to an electromagnetic field and the the Meissner effect are perfectly well calculated. They did not calculate the longitudinal response, which, in the BCS mean-field theory cannot be correctly calculated. This should not have been much of an issue but was made into one. In a system of charged fermions, one must couple the phase degree of freedom, whose gradient is proportional to the charge-current, to the electromagnetic field. The longitudinal charge response of the superconductor is then essentially the same as that of a normal metal~\cite{Anderson1958, Nambu1960}, i.e., at long wavelength the superconductor must also have plasmon excitations. Deriving the longitudinal response requires a theory respecting the equation of continuity of charge. The simplest such theory is not the BCS mean-field theory but the random phase approximation, or, in field theory parlance, the one-loop approximation. 

Much attention was paid in the years immediately following the BCS to the fluctuations of the amplitude of the order parameter near $T_c$, where they are 
over-damped and mixed with the fluctuations of the phase. Not much attention was paid to their properties at low temperatures. For the state of the theory of collective modes in 1969, see Ref. (\onlinecite{Martin1969}). Thus, for example, although time-dependent equations for the order parameter (containing  both first- and second-order derivative terms in time) were derived~\cite{Abrahams1966, Caroli1967} and the dispersion of the collective modes was calculated, the derivation, as pointed out by the authors~\cite{Abrahams1966}  themselves, was within limits where the answers were not valid. Therefore, the correct dispersion for the amplitude modes at low temperatures was not obtained. A simple calculation in the collisionless limit~\cite{Littlewood1981} at low temperatures yields the correct result for the amplitude modes. Perhaps more subtle than that, as discussed below, is how one couples to it in experiments.

Meanwhile, in the early 1960s, locally gauge-invariant gauge theories for matter fields had become popular in particle physics ~\cite{Hoddeson1997} because they gave answers that were well controlled, i.e., the theory was renormalizable. 
However, such theories predicted, as in superfluid He or weakly interacting Bose gas~\cite{Bogoliubov1947,*Bogoliubov1947b},  a massless phase fluctuation mode in the limit of long wavelengths. No massless particles were found in nature except the photons. The derivation of such massless modes by Goldstone~\cite{Goldstone1961, *Goldstone1962} and collaborators had left a little loophole in the existence of such a mode in the $\text{U}(1)$ models and in their generalization to non-Abelian models.  In 1963, in a generally overlooked paper, possibly because the discussion was not Lorentz-invariant, Anderson~\cite{Anderson1963} pointed out that the remedy in particle physics may be the coupling of the matter field to electromagnetism, the same remedy that converts the Goldstone modes in a neutral system to the plasmons in superconductors. Meanwhile, Nambu~\cite{Nambu1960}, who was the pioneer in connecting what was learned in the theory of superconductivity to particle physics and introducing the concepts of broken symmetry to particle physics, together with Jona-Lasinio~\cite{Nambu1961} applied the pairing idea to strong interaction physics, but not to weak-interaction physics and without cojoining to electromagnetism.

In 1964, three papers appeared~\cite{Higgs1964, Englert1964, Guralnik1964} on the issue of coupling of the matter field to the electromagnetic field and the problem of the massless mode. Of these three, the simplest is the paper by Higgs, which starts by writing down just the 
$\text{U}(1)$ Lagrangian coupled to the electromagnetic field, i.e., the Ginzburg-Landau type Lagrangian above Equation~\eqref{action}, and acknowledging that his work is based on the developments by Anderson and others in superconductivity. But being a particle theorist, he considered a  Lorentz invariant or particle-hole symmetric version of it and so put $K_1 = 0$. One then has an equation of motion with a second order time-derivative.  Along with the result that the Goldstone mode moves to a finite frequency,  he noted the existence of another mode with the words, ``It is worth noting that an essential feature of this type of theory which has been described in this note is the prediction of an incomplete multiplet of scalar and vector bosons''. The scalar is massive. The ``Higgs mode'' was born through this important remark. In particle physics, these issues took  on  
great importance  when Weinberg~\cite{Weinberg1967} and Salam~\cite{Salam1964,*Salam1969} used these ideas in connection with a gauge-invariant theory of weak interactions that had to be of $\text{SU}(2)$ symmetry. This was physics with application to observable particles and several predictions. In fact, the common textbook version of electro-weak theory is all found in the brief elegant paper by Weinberg~\cite{Weinberg1967} .
The vector bosons, W and Z  were derived; W bosons are the ideological equivalents of the plasmons, and Z bosons are the associated neutral oscillations in $\text{SU}(2)$ fields. They were observed in experiments in the 1970s. The massive scalar particle, the ``Higgs boson," appears to have been observed recently at the Large Hadron Collider (LHC)~\cite{CMS2012,ATLAS2012}, completing  the observations of the principal predictions of standard model and making us believe that the universe is, among other things, an electro-weak condensate. 

The phenomenological Standard Model assumes an electro-weak condensate. $\Psi({\bf r}, t)$ is often called the Higgs field. The equivalent of what is the superconducting condensate in electro-weak theory, is called the ``Higgs condensate". The excitations of this condensate are the Higgs modes and the vector bosons W and Z. The Higgs modes are the amplitude modes of this condensate, and the W and the Z particles are the $\text{SU}(2)$ phase modes.

The microscopic theory for the condensate, if it exists, is  yet to be discovered. To a condensed matter theorist, the principal difficulty, experimentally and notionally, is that the equivalent in electro-weak theory of what condensed matter physicists call the ``normal state," whose understanding was essential to understanding superconductivity, is estimated to occur at temperatures of over $250\,\text{GeV}$. No controlled experiments or observations are expected at such temperatures. In their absence, speculative directions of theory have prospered, which may wither in the absence of experimental support~\cite{Shifman2012}. It also appears that the analog of the BCS microscopic theory, i.e., condensation of fermions to form the superconducting condensate, the so-called ``techni-color" theories, are not supported in experiments. New ideas for theory and experiments are necessary. Perhaps detailed experimental results from the LHC in the future will help frame the right questions. 

The standard model, in its present form, would have the Higgs field existing above the transition temperature with zero vacuum expectation value, because 
$r < 0$ in the action Equation~\eqref{action}. In that case, only the leptons with weak and electromagnetic interactions, such as the electrons, are expected to become massless. An amplitude mode at finite energy would continue to exist, but the absence of the vacuum expectation value for $\Psi$ makes the $W$ and the $Z$ particles massless. All of this is quite different from the theory of normal metals. 

Motivated by some unexpected experimental observations~\cite{Sooryakumar1980,*Sooryakumar1981}, the theory of  ``what particle-physicists would call a Higgs mode", (to quote Higgs himself~\cite{Higgs1997}), in superconductors was discovered in 1981~\cite{Littlewood1981, *Littlewood1982}. In ignorance of developments in particle physics, it was called the amplitude mode. It was also pointed out that modes in other $\text{U}(1)$ systems, such as some antiferromagnets and incommensurate charge density waves (CDWs) can be similarly discussed.  The realization that cold bosons in a lattice have a line in the chemical potential-dimensionless interaction plane, which is particle-hole symmetric~\cite{Huber2007}  has led to the search for and evidence of the Higgs mode in such a system. The x-y antiferromagnetic insulators are particle-hole symmetric, and their modes naturally map to those of the action Equation~\eqref{action}. Specifically, a recent study of antiferromagnets has identified the modes corresponding to fluctuations of the magnitude of the order parameter as the Higgs modes, whereas the usual spin waves are the equivalent of the phase modes~\cite{Ruegg2008}.

One might well ask when is it appropriate to call an amplitude mode in condensed matter physics a Higgs mode. Higgs was dealing with a locally gauge-invariant problem. Of the condensed matter problems discussed here, only the superconductors are locally gauge invariant, the cold bosons in a lattice are only globally gauge invariant, while the antiferromagnets are even locally anisotropic. So it might be appropriate to refer to only the amplitude mode(s) in superconductors as Higgs mode(s). This semantic distinction is, however, not being maintained in the literature, and it may not be possible to stem the tide.

\section{Phenomenology}
\label{sec:phenomenology}
In subsequent sections, we discuss that, for (neutral) superconductors or for bosons on a lattice or for planar antiferromagnets,  the microscopic physics leads to the action of Equation~\eqref{action}, in the appropriate continuum limit. 
The equations of motion for fluctuations of $\Psi({\bf r}, t)$ are given from Equation~\eqref{action} by setting $\partial S/\partial \Psi^* =0$ and $\partial S/\partial \Psi =0$. The low-energy modes are obtained by expanding around the static equilibrium (with equilibrium $\phi$ chosen to be $0$) and separating into amplitude and phase fluctuations, $\delta_a$ and $\delta_{ph}$, respectively:
\be
\Psi({\bf r}, t) - \Psi_0 \approx \delta \Psi({\bf r},t) + i \Psi_0 \phi({\bf r},t) +... \equiv  \delta_a({\bf r},t) + i\delta_{ph}({\bf r},t) + ...
\label{eq:cartesian}
\ee
On Fourier transforming to wave-vector $q$ and frequency variables $\omega$,
\begin{subequations}
\label{phenEqs}
\begin{align}
(2r  + \xi^2 q^2 - K_2 \omega^2)\,\delta_a + i K_1 \omega \, \delta_{ph}= 0; \\
-iK_1\omega \,\delta_a + ( \xi^2 q^2 - K_2 \omega^2)\,\delta_{ph} = 0.
\end{align}
\end{subequations}
Some important conclusions~\cite{Varma2002} follow from these simple equations. 
First consider the particle-hole {\it asymmetric} case, with $K_2 =0$.
The roots of the secular equation are simply the degenerate Bogoliubov modes with dispersion $\omega^2 = \Big(\frac{2r}{K_1^2 }\Big) (\xi q)^2$ at long wavelengths. The eigenvectors show that the phase $\delta_{ph}$ and the amplitude $\delta_a$ are coupled so that  there are no distinct modes for the amplitude and phase fluctuations. 
\begin{figure}
\includegraphics[width=0.99\textwidth]{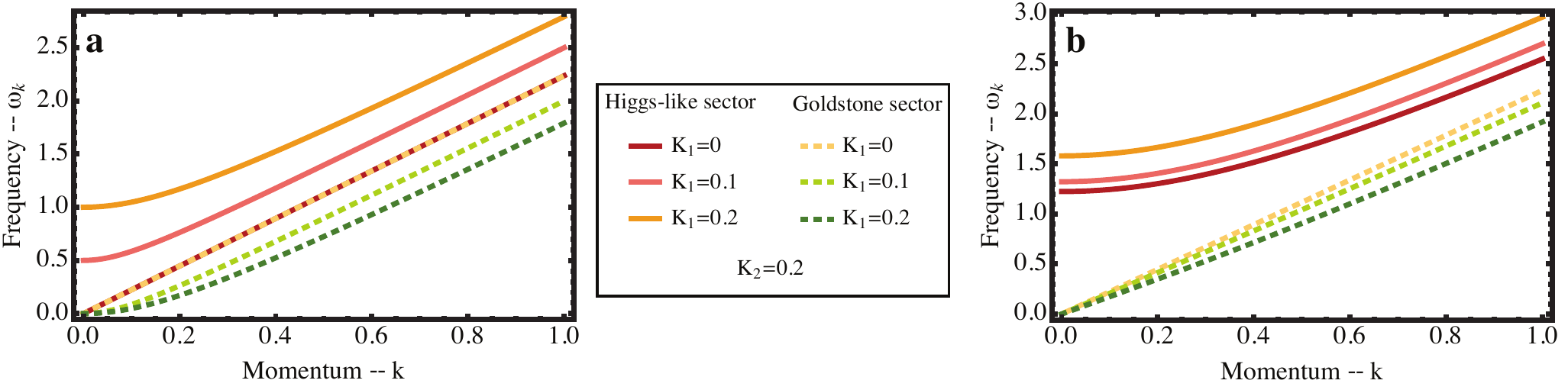}
\caption{{\bf (a)} The dispersion of the Higgs-like and the Goldstone modes at the critical point $r=0$ for several values of $K_1$ (as indicated in the legend). {\bf (b)} The dispersion in the superfluid (ordered) phase $r=0.15$, values of $K_1$ same as in \bf{a}. }
\label{fig:modes}
\end{figure}
Consider now the completely particle-hole symmetric by putting $K_1 = 0$, $K_2 \ne 0$. We get two orthogonal modes with eigenvector $\delta_{ph}$, i.e., the phase or Bogoliubov mode with dispersion 
$\omega^2 = (\xi q)^2/K_2$ and a mode with eigenvector $\delta_a$, i.e., the amplitude mode, with energy
$\omega^2 = 2r/K_2$ at $q =0$.
A few solutions for arbitrary ratios $K_1/K_2$ are shown in Figure~\ref{fig:modes}. The magnitudes of $K_1$ and $K_2$ are easily determined microscopically in weak-coupling calculations, which are the only analytic calculations available so far. They are, however, subject to renormalization for stronger interactions~\cite{Sauls2013}, an investigation that ought to be pursued further.

The important point is that having a Mexican hat potential is not enough to get distinct amplitude and phase modes. In particle physics, the imposition of Lorentz invariance guarantees $K_1 = 0$. To get orthogonal amplitude and phase modes in condensed matter physics, exact particle-hole symmetry is required. [A small amount of particle-hole asymmetry results in the mixing of the amplitude and phase modes. As a result of this mixing, the Higgs-like mode is pushed to higher frequencies and remains a gapped ($\omega_{q=0}=\frac{1}{K_2}\sqrt{2 r+K_1^2}$) even at the transition point.]  Normal metals are not particle-hole symmetric, but as explained below, superconductors are particle-hole symmetric at low energies of the order of the superconducting gap. In homogeneous superfluid $^4$He, one may again dispense with damping for low temperature, but there is no particle-hole symmetry: The time dependence is  nonrelativistic, as in Schr\"{o}dinger equation or the Gross-Pitaevskii equation. In cold bosons in lattices, as explained below, particle-hole symmetry may be carefully organized along certain lines in the chemical potential-interaction phase diagram. In dimer-antiferromagnets, the particle-hole symmetry corresponds to the equivalence of the two triplets that describe the easy-plane.

We now briefly mention the effect of an applied electromagnetic vector potential ${\bf A}({\bf r}, t)$ in superconductors through change of the gradient terms in Equation~\eqref{action}. Additional off-diagonal terms $\propto i e q A$ then appear in Equation~\eqref{phenEqs}. For particle-hole symmetry, $K_1 =0$, there is then a nonlinear in ${\bf A}$ driving of the mode and a modification of its dispersion.  For $K_1 \ne 0$, there is a linear response of $O(q K_1 A)$, because the amplitude and the phase modes are already linearly coupled by $K_1$.

\begin{figure}
\centerline{\includegraphics[width=0.9\textwidth]{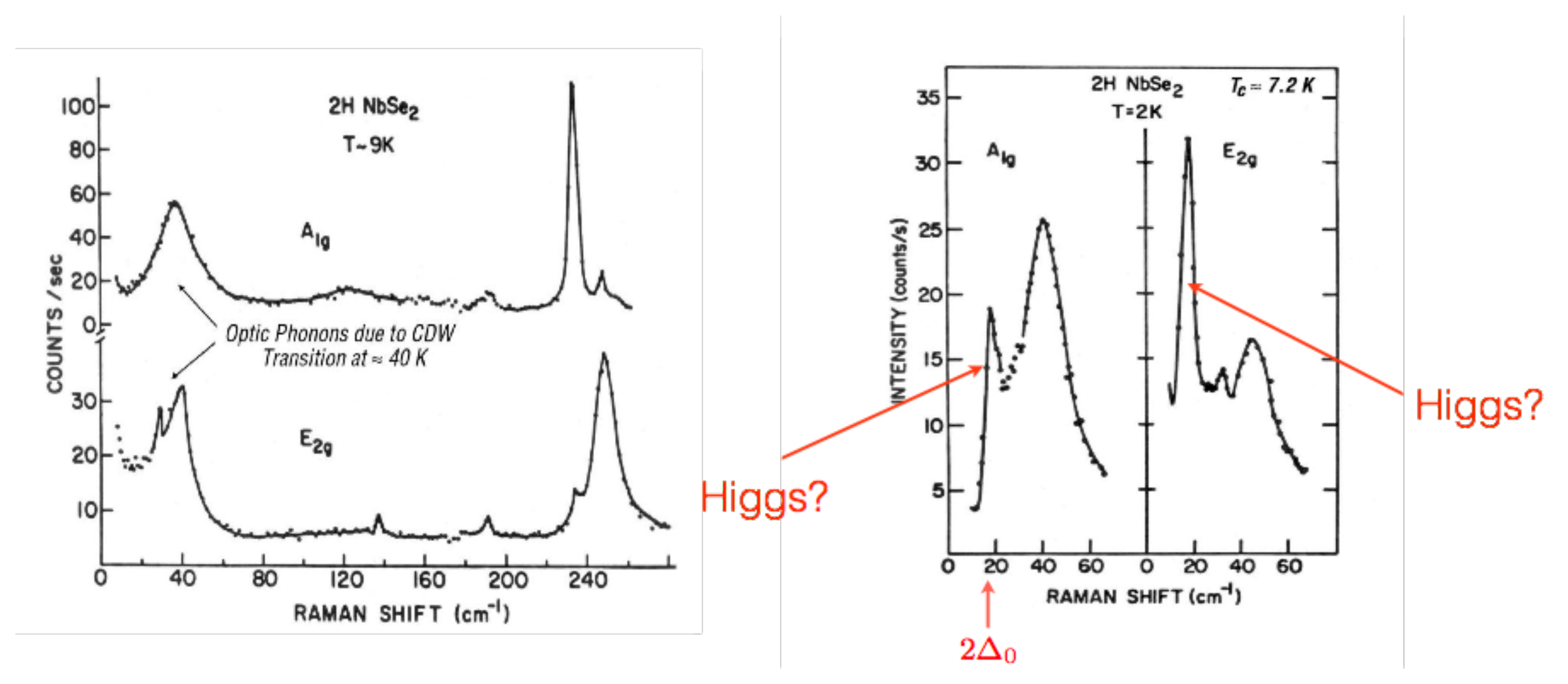}}
\caption{The Raman spectra in NbSe2 in two different symmetries $A_{1g}$ and $E_{2g}$. The Figure on the left shows the Raman spectra at 9 K, the superconducting transition temperature is 7.2 K. There is a charge-density wave transition in the compound at about 40 K. The low energy peaks near 40 cm$^{-1}$ occur only below 40 K and are due to the lowered lattice symmetry. The figure on the right shows the low-frequency part of the Raman spectra at 2 K, exhibiting the new modes near an energy of $2\Delta$. Adapted from Reference~\onlinecite{Sooryakumar1980,*Sooryakumar1981}.}
\label{Raman_Klein}
\end{figure}

\section{Microscopic theory in Superconductors}
In 1981, Raman scattering experiments were reported in NbSe$_2$ (see Figure~\ref{Raman_Klein}), both in its normal state and in its superconducting state, which occurs below $T_c \approx $ 7.2 K~\cite{Sooryakumar1980,*Sooryakumar1981}. Generally, Raman experiments are hard  to do on metals. The experiment could be done in the superconducting state only below the temperature of the superfluid temperature of $^4$He -- used as the coolant -- about 2 K. The Raman spectra (see Figure~\ref{Raman_Klein}) show optical phonons, which are also seen above the superconducting transition, and an additional peak at about twice the superconducting gap, $2\Delta$.  A microscopic theory to explain this observation was developed, and it was called the amplitude mode. It was realized that such a mode in principle exists in all superconductors and in any condensed matter problem with $\text{U}(1)$ symmetry but that, except under special circumstances, it may be over-damped. More importantly, the mode is a scalar, to which there is no linear coupling to the usual condensed matter probes. Therefore, it can be observed only in special circumstances and with special experimental tools. 

The key prediction of the theory, that the mode can only be discovered in Raman by stealing weight from some other excitation that shakes the superconducting condensate, has been verified only recently~\cite{Measson2014}. 

The microscopic theory of the Higgs or amplitude collective mode in superconductors is best formulated in the basis of the Gor'kov spinors: 
\be
\label{Gorkov}
\Phi_{\bf k, \alpha} = \begin{pmatrix}
      c_{\bf k \uparrow}    \\
      c^{\dagger}_{\bf -k \downarrow}  
\end{pmatrix}; ~~~
\Phi^{\dagger}_{\bf k, \alpha} = \begin{pmatrix}
c^{\dagger}_{\bf k, \uparrow} ~~
      c_{\bf -k, \downarrow}\end{pmatrix};~~~  
\Phi_{\bf k, \beta} = \begin{pmatrix}
      c_{\bf k \downarrow}    \\
      c^{\dagger}_{\bf -k \uparrow} \end{pmatrix}; ~~~
\Phi^{\dagger}_{\bf k, \beta} = \begin{pmatrix}
c^{\dagger}_{\bf k, \downarrow} ~~ c_{-\bf k \uparrow}
   \end{pmatrix}.
\ee
$ c_{{\bf k} \uparrow}, c^{\dagger}_{\bf k, \uparrow}$ are fermion annihilation and creation operators with momenta ${\bf k}$ and the z-component of spin. In this space, the charge-density operators appear as the $\tau_3$ component of the Pauli matrix ${\bf \tau}$, and the superconducting amplitude and phase as the $\tau_1$ 
and the $\tau_2$ components, respectively. $\alpha \ne \beta$ are up and down spin-components in some chosen quantization axis. Condensed matter physics is concerned with
only the interaction between charges so that the basic Hamiltonian is purely in the $\tau_3$ space:
\be
\label{Hfull}
{\cal H} = \sum_{\bf k, \alpha} \Phi^{\dagger}_{\bf k, \alpha} \epsilon_{\bf k} \tau_3 \Phi_{\bf k, \alpha} 
 + \sum_{\bf k, k',q, \alpha, \beta} V({\bf k,k',q}) \Phi^{\dagger}_{\bf k+q, \alpha}\tau_3 \Phi_{\bf k+q, \alpha}\Phi^{\dagger}_{\bf k'-q, \beta}\tau_3 \Phi_{\bf k', \beta}
 \ee
For attractive interactions, the normal Fermi-liquid state of such a Hamiltonian is unstable to Cooper pairing and superconductivity, and the ground state is given by the groundstate of the BCS Hamiltonian. It is useful to rewrite the Hamiltonian as the BCS mean-field Hamiltonian and its difference from ${\cal H}$:
\be
\label{bcs}
{\cal H} &=& {\cal H}_{BCS} + {\cal H}_1 \\
{\cal H}_{BCS} &=& \sum_{\bf k, \alpha} \Phi^{\dagger}_{\bf k, \alpha} \big(\epsilon_{\bf k} \tau_3  + \Delta_{\bf k} \tau_1\big)\Phi_{\bf k, \alpha} . 
\ee
As pointed out by Nambu~\cite{Nambu1960}, the BCS Hamiltonian bears a one-to-one correspondence with the Dirac Hamitonian. It is particle-hole symmetric, and equations of motion derived from it must be Lorentz invariant. This is strictly true only if the density of states in the vicinity of the chemical potential has a zero derivative in the normal state. If the density of states in the normal state of a metal has a finite derivative $\rho'(\epsilon)$ near the chemical potential, it is usually of $O(\rho(\epsilon_F)/W)$, where $W$ is the bandwidth. Then the particle-hole asymmetry in the quasiparticle density of states in the superconductor up to energies of several times the gap $\Delta$ is of $O(\Delta/W)$, which is typically $10^{-3}$. 
\subsection{Phase Modes}
In Equation~\eqref{bcs}, the phase of ${\Phi}$ has been arbitrarily chosen in the $\tau_2$-direction. Identical physical results should be obtained for any choice in the $\tau_1-\tau_2$ plane, i.e., by the rotation 
\be
\label{gauge}
\Phi({\bf r}, t) \to e^{i\phi({\bf r},t)\tau_3} \Phi({\bf r}, t).
\ee
This is a {\it local} gauge transformation under which the original Hamiltonian Equation~\eqref{Hfull} is invariant provided we also
make the change 
\be
\nabla \to \nabla + i \phi({\bf r}, t) \tau_3.
\ee
This invariance reflects the continuity equation for charge
\be
\label{cont}
\frac{\partial}{\partial t}\left (\Phi^{\dagger} \tau_3 \Phi\right) + \nabla \cdot \left(\Phi^{\dagger} \frac{{\bf P}}{m}\tau_3 \Phi\right)
\ee
where ${\bf P}$ is the momentum operator.

Although the full Hamiltonian satisfies this invariance, ${\cal H}_{BCS}$ does not -- the gauge invariance trouble mentioned above. The 
remedy is to calculate in approximations consistent with the invariance, the simplest being the random phase approximation (RPA). This requires fluctuations over the BCS mean-field Hamiltonian, with use of  ${\cal H}_1$ and writing down equations of motion for interaction vertex in the $\tau_3$ channel. Solving the equations of motion in the RPA approximation, one finds that the vertex has a pole at 
$\omega = (v_F/3)k$ for $k \to 0$ if one neglects the  Coulomb interaction part of ${\cal H}_1$.  This would be the Goldstone or the Bogoliubov mode, whose physics we have already interpreted in the discussion of Figure~\ref{Fig:mexhat}. But unlike in the case of a neutral superfluid like $^4$He, this is a mode of oscillation of charged particles. The Coulomb interaction part of ${\bf H}_1$, which is singular as $q^{-2}$, cannot be neglected. Inclusion of this process pushes the pole at long wavelengths to near the plasma frequency ${\Omega_p} = (4\pi n e^2/m)^{1/2}$~\cite{Schrieffer1964} with negligible corrections of order $(\Delta^2/E_F)$. Here,
$N$ is the charge density of the metal and $E_F$, its Fermi energy. $\Omega_p$ in a typical metal is on order of several eV's, whereas $\Delta$ is typically of O(1 meV). So, as far as the longitudinal properties are concerned, essentially nothing changes on the transition from the metallic state to the superconducting state. It would be very surprising if it did. After all, the color of a metal is determined to a large extent by the plasma frequency; one would hardly expect it to change due to the changes of the excitations at tiny energy compared with the plasma frequency due to superconductivity. This coupling of matter field to electromagnetic field, however, has been of great importance~\cite{Hoddeson1997} in high energy physics, where one does not have a given microscopic Hamiltonian and where there are no experiments in the ``normal state".
 
\subsection{Amplitude Mode}

To find the amplitude collective modes, one must look for singularities in the vertex in the $\tau_1$ channel. The bare vertex in the $\tau_1$ channel is just $\Delta$, as may be seen from the BCS Hamiltonian (Equation~\eqref{bcs}). The simplest next step is again RPA. Just as RPA in the density channel can be  trusted because it is consistent with the invariance of the full Hamiltonian to the transformation (Equation~\eqref{gauge}), there are additional but rather unfamiliar invariances of the full Hamiltonian (Equation~\eqref{Hfull}) in the $\tau_1$ and $\tau_2$ channels. These were pointed out by Nambu~\cite{Nambu1960} in Eqsuations (4.4) - (4.8) in his classic paper, where the continuity equations in the Cooper channels $\tau_1, \tau_2$ were provided.

\begin{figure}
\centerline{\includegraphics[width=0.6\textwidth]{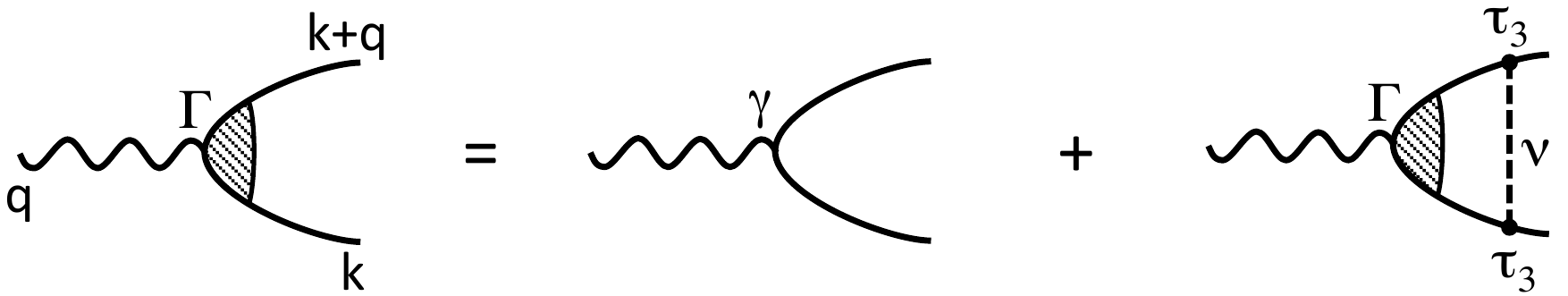}}
\caption{The vertex equation or the Bethe-Salpeter Equation for the amplitude mode or the Higgs. The fermion propagators in this diagram are in the representation of the Gor'kov spinors, Equation~\eqref{Gorkov}. The bare-vertex $\gamma$ as well as the renormalized vertex $\Gamma$ are in the $\tau_1$ channel. Projected to the ordinary fermion operators, this diagram is in the particle-particle or Cooper channel with net momentum ${\bf q}$ and spin-singlet.}
\label{Fig:vert-tau1}
\end{figure}

The RPA equation of motion for the vertex, which is consistent with the Nambu continuity equations in the Cooper channel, is represented graphically in Figure~\ref{Fig:vert-tau1}. Writing $\Gamma = \zeta(q)\tau_1$, the collective modes are given by the solutions of the homogeneous part of the equation for 
Figure~\ref{Fig:vert-tau1}, i.e., by setting the value of the bare vertex $\gamma =0$, which by inspection of this figure is
\be 
\label{homoeq}
\zeta(q) \tau_1 = -V \int \frac{d^4k}{(2\pi)^4}\tau_3 G(k+q) \zeta(q) \tau_1 G(k) \tau_3. 
\ee
Her,e $k$ stands for the four-vector for the energy-momentum carried by the propagators. The solution of Equation~\eqref{homoeq} is described in detail in Refs.~\onlinecite{Littlewood1981, *Littlewood1982}. For $T=0$, at momentum transfer $q \to 0$, it 
 has singularities ~\cite{Littlewood1981, *Littlewood1982} at  the energy $\nu$  given by
\be
\label{ampvertsoln}
1 + V\sum_{\bf k} \frac{\epsilon_{\bf k}^2}{E_{\bf k}(\nu^2/4 -E^2_{\bf k})} = 0.
\ee
V is the s-wave attractive interactions between normal-state particles, which are also shown in Figure~\ref{Fig:vert-tau1}); $E^2_{\bf k} = \epsilon^2_{\bf k} + \Delta^2$ is the BCS quasi-particle energy assuming the simplest s-wave, ${\bf k}$-independent gap function $\Delta$.  
As usual, singularities in the vertex signify collective modes. Several important points about Equation~\eqref{ampvertsoln} should be noted. At $\nu = 2 \Delta$, this is just the BCS equation for the superconducting gap. The finite temperature version of this equation has a factor $\tanh(\beta E_{\bf k})$ in the numerator in the second term. At $\nu \to 0, \Delta \to 0$, this is then  just the BCS condition for the temperature of the superconducting instability.  This means that every (s-wave) superconductor is required to have a collective mode in the $\tau_1$ or amplitude channel at least in the weak coupling limit and for the momentum $q \to 0$ and provided particle-hole asymmetry is negligible. Comparing this with the phenomenology found in Section II and Figure~\ref{Fig:mexhat}, this is the same collective excitation as the amplitude or Higgs mode of the phenomenological $\text{U}(1)$ model. The physical reason for getting the collective exception in the weak-coupling limit at precisely $\nu = 2\Delta$ and the relation to the phenomenology is provided by taking the coefficients $r$ and $U$ in Equation~\ref{action} from BCS theory and realizing that the coefficient $K_2$ is given by the compressibility, which is $N(0)$, the density of states at the Fermi surface.

To consider the observability of such modes, one must however also consider the damping of such resonances.
With the benefit of a microscopic theory, we can calculate the finite $q$-correction and the effect of quasiparticle lifetime. Solutions of Equation~\ref{homoeq} at small ${\bf q}$~\cite{Littlewood1982} gives 
\be
\label{nuq}
\nu^2(q) = 4\Delta^2 + \frac{1}{3} v_F^2q^2 + i \frac{\pi^2}{24}(v_Fq)\Delta,
\ee
where $v_F$ is the Fermi-velocity. The damping at finite $q$ is because the frequency of the mode lies in the particle-hole continuum. We see below that 
the mode is shifted below $2 \Delta$ because of interactions with phonons, and then there is then no damping untill a minimal value of $q$. There are bound to be corrections to this result due both to Fermi-liquid effects and strong-coupling, which have not been much pursued.

An important question is what do Coulomb interactions do to the amplitude mode. The mode is an eigenvector of $\tau_1$. It is therefore orthogonal to the charge sector $\tau_3$. It can therefore not couple to Coulomb interactions. This is seen in explicit calculations~\cite{Littlewood1981, *Littlewood1982} but only for particle-hole symmetry. This may also be seen from the fact that $\Phi^{\dagger} \tau_3 \Phi$ is invariant to the transformation Equation~\eqref{gauge}.  

\subsection{Coupling of Amplitude Modes to Experimental Probes.}
If one puts in the Raman scattering $q$, which is of the order of the inverse wavelength of light in Equation~\eqref{nuq}, the damping of the mode is much larger than $2\Delta$. This bears no relation to the experiments
that see very sharp peaks. But even more important is the question how one may couple with external fields to such a mode and observe it as a linear response. As we have seen, the Higgs mode has no projection to charge or current density or dipole or magnetization oscillations and therefore no linear coupling to electromagnetic fields. Indeed, it has no linear-response coupling to ordinary condensed matter probes. This corresponds to the oft-used characterization of the Higgs as a scalar. 

\begin{figure}
\centerline{\includegraphics[width=0.85\textwidth]{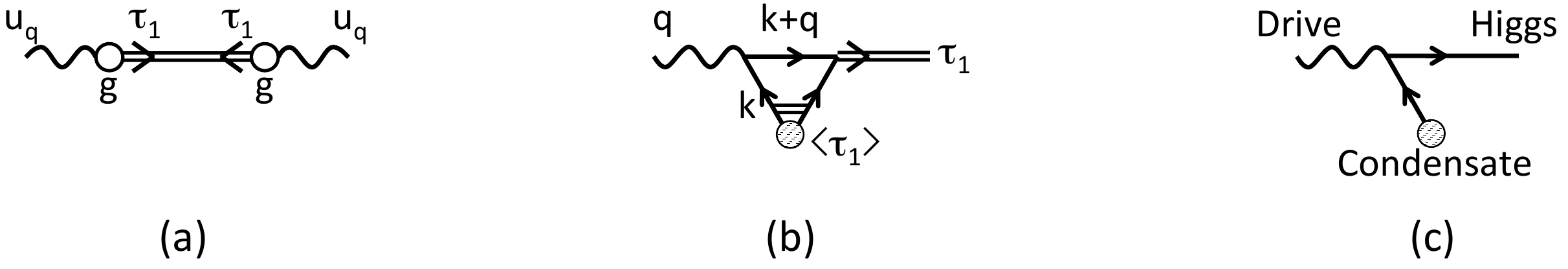}}
\caption{(a) The Raman active phonon's self-energy due to linear coupling to the Higgs mode due to the anomalous vertex of Equation~\ref{couple}. A similar process gives the self-energy of the Higgs mode with an intermediate state of the phonon.  (b) The anomalous vertex which relies on the existence of a number non-conserving condensate. (c) Vertex for production of Higgs boson, which may be used in cold boson atoms in a lattice, similar to that of (b). The drive excites long wave-length density or current fluctuations which are quadratic in boson operators. A vacuum expectation value of the boson operator exists in the condensed state leaving the fluctuations of the other to turn into the Amplitude or Higgs mode. }
\label{coupling}
\end{figure}

A reading~\cite{Littlewood1981} of the experiments~\cite{Sooryakumar1980,*Sooryakumar1981} suggested that the new mode appears only through stealing spectral weight from the optical phonon present already in the Raman spectra above $T_c$. (Approximate conservation of the first moment of the spectral weight in the two modes as a function of a magnetic field was shown in the experiments, but one finds an equally good conservation of the integrated spectral weight itself in those measurements. No temperature dependent measurements in the superconducting state were made until very recently~\cite{Measson2014}.) Such a conservation automatically occurs if the Higgs were {\bf linearly} coupled to Raman active optical phonons.  The Higgs mode would then appear as a pole in its self-energy of the Raman active phonon, resulting in a pair of resonances, with the weight of the two resonances the same as that of the only resonance above the superconducting transition temperature to which alone the external optical probe couples.

It was therefore natural to assume that a coupling of the lattice displacement coordinate $u_{\lambda, q}$ of the Raman active modes to the amplitude mode $(\Psi^{\dagger}\tau_1 \Psi\big)({\bf q})$ exists:
\be
\label{couple}
H' = \sum_{\lambda, {\bf q}} ~g_{\lambda} ~u_{\lambda, {\bf q}} ~(\Psi^{\dagger}\tau_1 \Psi\big)({-\bf q}) + H.C.
\ee
Here, $\lambda$ labels the polarization of the displacement coordinate. Now we can calculate the self-energy of the displacement propagator,
$D(\lambda, {\bf q}, \omega) \equiv \langle u_{\lambda, q}u^{\dagger}_{\lambda, q} \rangle (\omega)$ as shown in part (a) of Figure~\ref{coupling}, as well as the self-energy of the amplitude mode due to the coupling. Such couplings have two effects: The pole of the amplitude mode shifts downwards at $q =0$ from its value without the coupling of $2\Delta$ if lies below the phonon energy $\omega_{\lambda,0}$. So at small $q$, it no longer overlays the particle-hole continuum and is therefore undamped. The renormalized pole of the amplitude mode appears in the self-energy of  $D(\lambda, {\bf q},\omega)$ to which external photons couple.  An approximately correct estimate of the spectral weight transfer in Figure~\ref{Raman_Klein} could be obtained for the  breathing or $A_{1g}$ mode through an estimate of the coupling $g_{\lambda}$ obtained from the relative variation of the superconducting and charge density wave transitions with hydrostatic pressure. 

The magnitude of the self-energy in Figure~\ref{coupling} depends on, beside the coupling matrix elements, the difference in the energy of the phonon $\omega_0$ and the amplitude mode, i.e., $2\Delta$. Calculations show that the effect disappears as $\frac{g^2_{\lambda}}{\omega_0}(\omega_0-2\Delta\big)^{-2}$. We have a fortunate situation in NbSe$_2$, where due to the occurrence of a charge density wave transitions at $T_{cdw} \approx 30 K$, there exist modes with $\omega_0 \approx 4 \Delta$. Other materials where such happy circumstances prevail are some A15 compounds in which the amplitude mode may also be identified through the conservation of total spectral weight~\cite{Hackl1983}.

It should be noted that Equation~\eqref{couple} appears to violate the conservation of particle number (not charge). To understand the conditions in which this is allowed, one has to delve a bit deeper into the physics of $g_{\lambda}$. The fact is that a lattice displacement or variation in (local) density couples only to the electronic charge density, i.e., to the $\tau_3$ channel. We can however generate a vertex of the form Equation~\eqref{couple} through the process shown in Figure~\ref{coupling}b.  As shown, we have a superconducting condensate that is a reservoir of a (even) number of particles or a state with a linear combination of different (even) number of particles. So the observation of the amplitude or Higgs mode occurs only through shaking the condensate by the lattice vibration and a particle-number nonconserving condensate. Similar freedom from electro-weak conservation laws is provided by the Higgs condensate in some of the processes by which Higgs excitations are discovered at the LHC.
Similarly for cold bosons in a lattice, one may observe the amplitude mode using the process shown in Figure~\ref{coupling}c, which, as we discuss in Section~\ref{sec:UCA} is yet to be employed directly.

Instead of the phenomenological coupling of Equation~\ref{couple}, there have been microscopic calculations~\cite{Levin1993, Kurihara1983} starting from the full panoply of equations for the superconductor and the charge density waves. In effect, the coupling vertex illustrated in the middle part of Figure~\ref{coupling} is calculated. These authors found that the general results given above are reproduced except for the fact that  instead of a constant coupling $g$, one has a frequency dependent coupling. These affect the detailed comparison with experiments. 

\begin{figure}
\centerline{\includegraphics[width=0.8\textwidth]{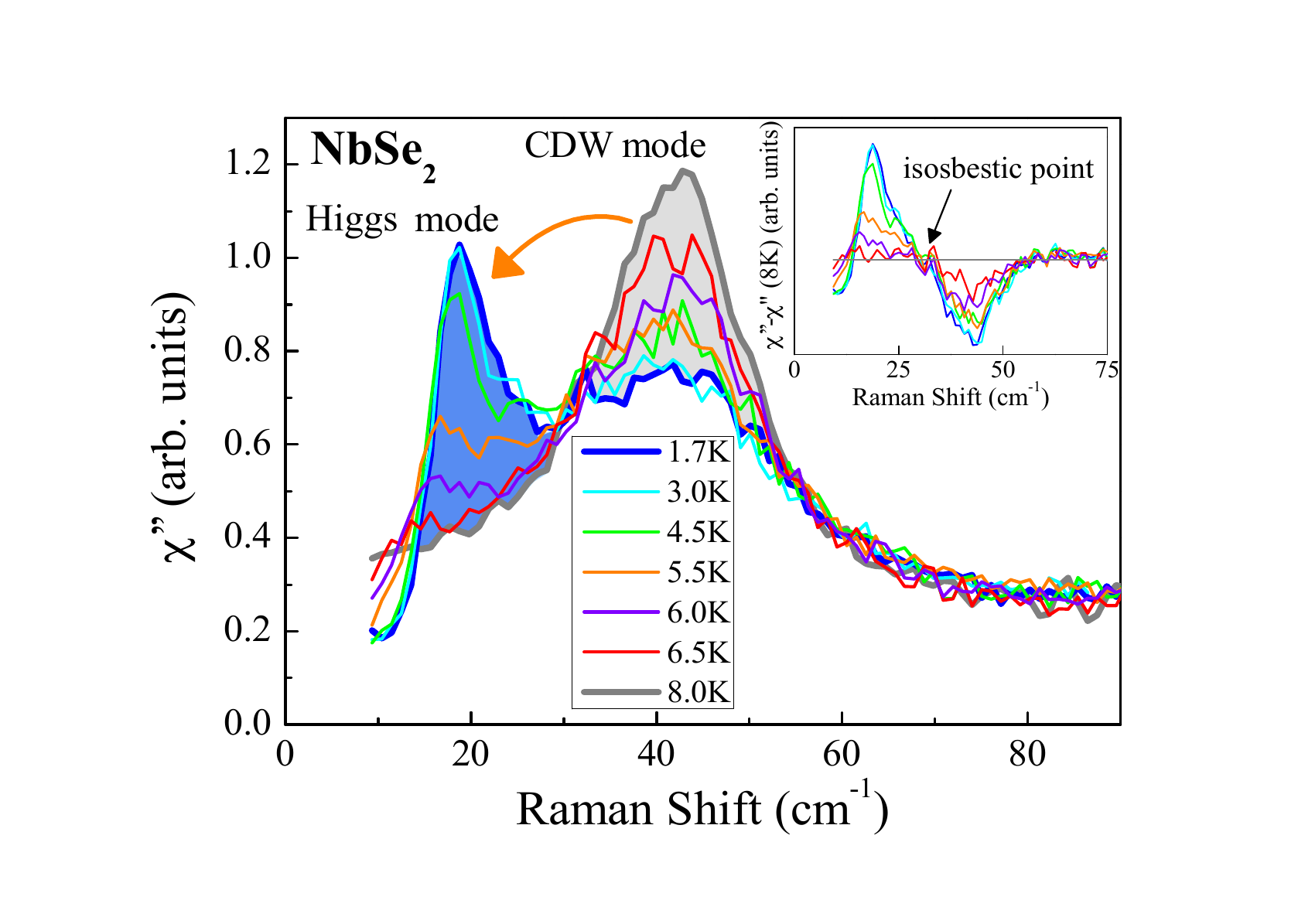}}
\caption{The  Raman spectra in the A$_{1g}$ symmetry at various temperatures in NbSe2, taken from Reference~\onlinecite{Measson2014}. The inset on the right shows the spectra at different temperatures below T$_c$ subtracted from the data just above T$_c$. Abbreviation: CDW, charge density wave.}
\label{Measson1}
\end{figure}

The evidence that observed excitation is the amplitude mode or Higgs is the evidence for it to be a scalar. In that case, it can have no direct linear coupling to the usual external probes. In a Raman experiment, it should appear only through stealing the spectral weight from some other excitation so that the spectral weight of the two resonances below $T_c$ is the same as that of the one resonance above $T_c$.  Measson et al.~\cite{Measson2014} were able to fdetect thid spectral weight transfer due to the improvement of techniques of performing Raman scattering experiments on metals for several temperatures from above $T_c$ to below $T_c$. The spectra obtained by them for various temperatures in NbSe$_2$ are shown in Figure~\ref{Measson1}, where in the inset the difference of the spectra at any temperature from the spectra just above $T_c$ is also shown. These experiments reveal a conservation of weight to within the experimental accuracy, which is approximately 5\%. 
In another important experiment, the same authors examined the Raman spectrum of NbS$_2$, a compound isostructural with NbSe$_2$ and thast has a similar T$_c = 6.2 \text{K}$. But this compound has no low-lying phonon because there is no CDW transition. No mode near $2\Delta$ appears in this case, consistent with the discussion above.

There is yet another revealing aspect of the earlier and the more recent experiments on NbSe$_2$. The charge density wave transition leads to low-energy phonons both in the breathing channel $A_{1g}$ and in a channel in a higher irreducible representation, $E_{2g}$.  A mode of almost the same energy, $2\Delta$ is observed below $T_c$ in both channels and with preservation of the spectral weight in each channel.  This is highly unusual and is more evidence that the vertex for coupling the amplitude mode is as shown in Figure~\ref{coupling}. Whichever channel, irrespective of its symmetry, shakes the condensate adequately and is at low enough energy to acquire enough amplitude of the singular self-energy by coupling to the amplitude mode helps to see the amplitude mode. 

One notices from Figure~\ref{Measson1} that the energy of the amplitude mode increases as the temperature is decreased. But this is not simply the temperature dependence of $2\Delta(T)$ of BCS theory, as in the simple theory that we have presented. There may be two reasons for the discrepancy. The mode energy is shifted from $2\Delta(T)$ due to coupling to the CDW phonon, which, as Figure~\ref{coupling} explains, depends on the coupling constant, which in turn depends on the magnitude of the condensate, which is temperature dependent.  Also NbSe$_2$ is a  strong-coupling superconductor that shows faster rise of $2\Delta(T)$ just below T$_c$ and a more uniform approach toward T$\to$ 0 than BCS theory. Both these factors correct from the temperature dependence of the BCS theory towards what is observed. However, there has been no attempt to calculate the temperature using these considerations to see whether there is a good fit with the recent experiments. In the microscopic calculation~\cite{Levin1993, Kurihara1983} of the coupling of the charge density wave to superconductivity, an estimate of the coupling constant $g_{\lambda}$ and its dependence on the frequency of the phonon is given. These have not been used to calculate the detailed temperature dependence because this the temperature dependence was only recently obtained. 

The occurrence of a resonance near 2$\Delta$ led to some theoretical efforts to understand it in terms of pair-breaking~\cite{Falicov1980, Dierker1984} in the $\tau_3$ or charge channel. The Hartree-Fock approximation using the BCS Hamiltonian~\cite{Falicov1980} gives such a resonance, but this is the old problem of calculating a non-gauge-invariant response in the longitudinal channel. Such a resonance occurs at the plasma frequency in the superconducting state, with energy almost equal to the plasma frequency of the normal metallic state in a proper theory that considers long-range Coulomb interactions~\cite{Littlewood1981,Littlewood1982}. It was then argued that because the intensity of the external radiation decreases from the surface of the sample over its penetration depth~\cite{Dierker1984}, there may be additional effects in the $\tau_3$ channel. However, a correct theory in the longitudinal channel due to effects of a surface can only produce surface plasmons in the superconductor (with some modifications at low energies). 
There has also been some attempt to explain the observed modes at energy 2$\Delta$ with the Bardasis-Schrieffer~\cite{Schrieffer1961} modes~\cite{Tutto1992}, which are modes of excited pair states in a symmetry channel other than that of the ground state). These can also be excluded by the observation of the conservation of weight and the occurrence of the mode of the same energy in different Raman symmetries.

There is also a more violent way of shaking the condensate to observe the amplitude mode in recent experiments. This is a pump probe experiment~\cite{Matsunaga2013} in which an intense THz electromagnetic pulse is applied to an Nb$_{1-x}$Ti$_x$N film below its superconducting transition temperature. The transient creation of a large density of particle-hole pairs shakes the condensate. The response of the condensate is then probed as a function of time, effectively by analogs of Raman scattering. It is found that a transient oscillation appears whose frequency is close to $2\Delta$. This process is closely related to other theoretical analysis~\cite{Barankov2006, Yuzbashyan2006}. It may also be regarded as the manifestation of the nonlinear coupling of the mode to electromagnetic fields, briefly discussed in Sec. II.

\subsection{Direct coupling to the condensate using the Josephson and proximity effects}

One can couple linearly and directly with external probes to the amplitude or Higgs modes using the Josephson effect. Consider  Josephson coupling between a superconductor and another of the same symmetry but, in general, different amplitude. The coupling energy between the two weakly coupled superconductors, one with order parameter $|\Psi| e^{i\phi}$ in which we are interested in exciting collective modes and the other with the order parameter $|\Psi_p| e^{i\phi_p}$, is
\be
\label{Joseph}
F_J = C |\Psi||\Psi_p| \cos(\delta \phi);\quad\quad\quad\quad \delta \phi = (\phi -\phi_p).
\ee
The coupling constant $C$ depends on the properties of the weak link. Add this to the action in Equation~\ref{action}. If a time-dependent voltage is applied across the two superconductors, the difference of phase evolved in time because it is canonically conjugate to charge difference. The derivative with respect to time of the phase difference is the chemical potential difference between the two superconductors or the applied voltage across them:
\be 
\frac{d\,\delta\phi}{dt} = eV(t).
\ee
If $V(t)$ is periodic with frequency $\omega$,  so is $\delta\phi(t)$. Now one gets a periodic driving term on the right side of both Equations~\eqref{phenEqs}. Therefore, both the phase mode and amplitude or Higgs mode can be excited by this method. This was the technique with which Carlson \& Goldman~\cite{Carlson1975} and Kadin \& Goldman~\cite{Kadin1982} observed a new class of phase oscillation modes (Carlson-Goldman modes) near $T_c$ as well as slightly below $T_c$ under conditions in which 
there is counterflow of supercurrent and normal current such that charge neutrality is maintained and the phase mode is not pushed to high energies. One can also apply a magnetic field parallel to the junction between the two superconductors which acts to introduce a spatially periodic coupling between the two superconductors along the surface of the junction orthogonal to the external magnetic field. One can then study the dispersion relations, $(\omega(k))$, of the collective modes.
It is straightforward to show that this process excites not only the Carlson-Goldman modes but the Higgs modes as well. In the experiments~\cite{Carlson1975}, there is a report of finding higher energy resonances, but this matter was not pursued.

\subsection{Higgs in superconductors of other symmetries}
Superfluid $^3$He has a multicomponent order parameter~\cite{Leggett} with angular momentum 1 and spin 1. The collective modes of oscillations between the various components of the order parameter have been studied by W\"{o}lfle~\cite{woelfle} and others~\cite{McKenzie2013, volovik2014}. 
In cuprate superconductors, which have a spin-singlet condensate with $d(x^2-y^2)$ lattice symmetry, there is a mysterious mode~\cite{sacuto} directly seen in Raman scattering in the $A_{1g}$ symmetry at an energy of approximately 1.5 $\Delta$, where $\Delta$ is the maximum gap as a function of angle on the Fermi surface. This appears below a continuum electronic scattering, which is characteristic of both the normal and the superconducting cuprates. It has been identified as the amplitude mode of the superconductor~\cite{barlas2014} which draws weight from the continuum, just as the Higgs draws its weight in Raman scattering from the low-lying optical phonons in NbSe$_2$.

\section{Higgs mode in ultra cold atom gases}
\label{sec:UCA}
In the past thirty years, a new field of experimental physics, that of ultracold atoms, has developed and made remarkable progress. Some of the technical highlights are the development of techniques for cooling atoms to low temperatures and subsequently trapping them (e.g., see~\onlinecite{Phillips1998}), evaporative cooling to form Bose-Einstein condensates~\cite{Davis1995}, the development of optical lattices that can be modulated and atomic imaging with single-site resolution,  and the observation of the superfluid-Mott insulator transitions~\cite{Greiner2002}. This progress has raised hopes of transforming atomic, molecular, and optical physics from a branch of physics that studied few atom interactions and thermal gases into a platform for studying many-body physics in quantum degenerate states~\cite{Jaksch2005}. 

This new setting offers some experimental advantages and some challenges compared with those in other condensed matter for studying collective modes of bosons. An important difference is that ultracold atom superfluids are charge neutral. So although there should be Higgs modes, there are also massless Bogoliubov modes. As distinct from superfluid $^4He$, ultra cold atom experiments are performed in optical lattices. The periodic lattice gives rise to particle-hole symmetry required for Higgs modes along a set of lines in the phase diagram. 

A unique aspect of the new setting is the ability to control Hamiltonian parameters over wide ranges, during the course of the experiment.  
This has allowed the observation of the superfluid-Mott insulator transitions~\cite{Greiner2002} which has been a necessary step in obtaining evidence for the Higgs mode in the superfluid. The present evidence for this mode is through the spectroscopy of absorption of energy of the atoms in response to periodic modulations of the kinetic energy (KE) and the correspondence of the observed line shape with analytic  as well as quantum Monte Carlo calculations. As we discuss, this is not as direct as the evidence for Higgs mode in superconductors. At the end of this section, we speculate on more direct observation of Higgs through further manipulations and the possibility of observing the Higgs of neutral fermions in their paired superfluid state.

\subsection{Ultracold atoms in optical lattices}
Consider a collection of repulsively interacting bosonic atoms in a potential created by the combination of an atom trap and an optical lattice. The motion of the atoms, in free space, is governed by the Hamiltonian
\begin{align}
H_{\text{FS}}=\sum_i \frac{p_i^2}{2 m} + \sum_i \tilde{V}(r_i) + \sum_{i,j} \tilde{U} \delta(r_i-r_j)
\end{align}
Here, $p_i$ and $r_i$ are the momentum and position of the $i$-th atom, $\tilde{V}(r)$ is the combination of the optical lattice potential and the trap potential for the atoms, and $\tilde{U}$ is the repulsive interatomic potential.  
The details of the physics of the trap potential and the periodic lattice potential in one, two, or three dimensions created by a standing wave of light may be found in Ref.~\onlinecite{Pethick2008}. 
The typical range of the interatomic interaction $\tilde{U}$ is significantly smaller than the lattice period, and hence it can be replaced by an on-site interaction. For some atoms a Feshbach resonance~\cite{Jaksch2005} can be used to tune the strength and sign of this on-site interaction. However for the case of the Higgs boson experiments, $^{87}$Rb atoms were used. Although the hyperfine levels of $^{87}$Rb atoms have a number of narrow Feshbach resonances, there is no convenient (i.e., wide) Feshbach resonance that can be used to tune the strength of interactions and so $\tilde{U}$ is fixed. 

\subsection{The Bose-Hubbard model and the Higgs excitation}
Consider the motion of the interacting atoms in a periodic potential. If the lattice potential is sufficiently deep,  the atoms are confined to the lowest Bloch band so that we can describe the low-energy sector of the system using the Bose-Hubbard model~\cite{Fisher1989}
\begin{align}
H_\text{BH}=-J \sum_{\langle i j \rangle} b_i^\dagger b_j + \frac{1}{2} U \sum_i n_i(n_i-1) -\mu \sum_i n_i.
\end{align}
Here, $b_i^\dagger$ and $b_i$ are the boson creation and annihilation operators on site $i$, $n_i$ is the occupancy of the $i$-th site. The hopping matrix element $J$ is related to the overlap of the boson wave functions on neighboring sites and the interaction strength $U$ to the boson scattering length in free space $a_s$. Both $J$ and $U$ are sensitive to the optical lattice depth; however in the case of a deep lattice, $J$ depends exponentially on the lattice depth, whereas $U$ varies weakly. The chemical potential $\mu$ is related to the depth of the trap and its filling, and generally varies across the system. However, we initially consider only the case of uniform $\mu$.

\begin{figure*}
\includegraphics[width=0.3\textwidth]{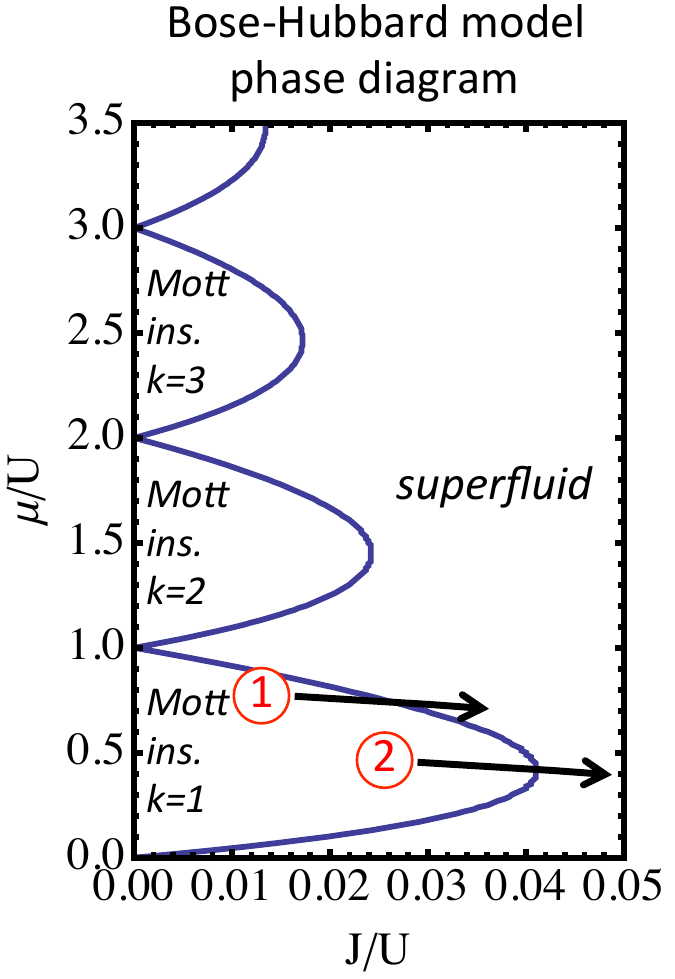}
\caption{
The phase diagram of the Bose Hubbard model in 2D. The phase diagram was obtained by optimizing the trial wave function Equation~\eqref{eq:PsiTrial} with up to 10 bosons on each site. Trajectories 1 and 2 refer to generic and particle-hole symmetric traversals of the phase diagram.
}
\label{fig:BoseHubbardPhaseDiagram}
\end{figure*}

To get a notion of the $T=0$ phase diagram of the Bose-Hubbard model consider a ground state wave function of the product form
\begin{align}
|\psi_{\text{prod}}\rangle=\prod_i \left(\alpha |0\rangle_i + \beta |1\rangle_i + \gamma |2 \rangle_i + \dots \right),
\label{eq:PsiTrial}
\end{align}
where the index $i$ runs over all the sites of the system, $\alpha$, $\beta$, $\gamma$, $\dots$ are complex coefficients, and the notation $|n\rangle_i$ indicates a state with $n$ bosons on site $i$. On minimizing $\langle \psi_{\text{prod}} |H_\text{BH}| \psi_{\text{prod}} \rangle$ with respect to $\alpha, \beta, ..$ two kinds of phases are found: Bose-Mott insulators and superfluids. The resulting phase diagram is plotted in Figure~\ref{fig:BoseHubbardPhaseDiagram}.

Deep in the Mott insulator phase, where $U \gg J$, the bosons are localized. In this case, the optimal trial wave function has exactly $m$ bosons per site $|\psi_{\text{prod}, m}\rangle=\prod_i |m\rangle_i$, where $m$ is determined by the value of the chemical potential $\mu$. Deep in the superfluid phase, $J \gg U$, the bosons delocalize and acquire phase coherence. In this case, the coefficients $\alpha$, $\beta$, $\gamma$, $\dots$ of the trial wave function are all finite, and hence the superfluid order parameter $\langle \psi_{\text{prod, SF}} | b_i | \psi_{\text{prod, SF}} \rangle = \alpha^* \beta + \beta^* \gamma + \cdots$ has a finite expectation value.  Although this simple wave function gets the layout of the phase diagram, it does not accurately predict the locations of the phase transitions. This is because of the absence of spatial correlations in the trial wave function. More accurate phase diagrams have been obtained using both more sophisticated theoretical treatments~\cite{Freericks1996,Dutta2012} and  numerical path integral quantum Monte Carlo simulations~\cite{Krauth1991, Capogrosso2007,Capogrosso2008}.  

The simplest method for describing the long wavelength modes of the Hubbard model in the vicinity of the Mott insulator to superfluid transition is to map the lattice model onto a continuum field theory (see \onlinecite{Sachdev2011}). This approach results in the action of the form in Equation~\eqref{action}. In the present case the tuning parameter that takes the system across the phase transition $r$ is proportional to $J/U-(J/U)_\text{critical}$. Negative $r$ corresponds to the symmetric phase, i.e., Mott insulator, and positive $r$ to the spontaneously broken symmetry phase, i.e., the superfluid state. 

As already explained, in the continuum description, particle-hole assymmetry depends on the value of $K_1$. Only for $K_1 = 0$ is the critical theory particle-hole symmetric or relativistic, and we see emergent Lorentz invariance. In the phase diagram of Figure~\ref{fig:BoseHubbardPhaseDiagram}, $K_1 =0$ is achieved along a line that meets the Mott-Insulator phases at the maximum of $J/U$ at each of the boundaries, where $\mu/U$ is half-integral.

To get a physical picture of the long wavelength ($k=0$) quasi-particle modes, we can extend the trial wave function of Equation~\eqref{eq:PsiTrial} by letting the coefficients $\alpha$, $\beta$, $\gamma$, $\dots$ be site dependent~\cite{Altman2002,Huber2007,Huber2008,Pekker2012}. The normal modes can then be obtained by extremizing $\langle \psi_{\text{prod}} | i \partial_t - H_\text{BH} |  \psi_{\text{prod}} \rangle$. 
First, let us look at the particle-hole symmetric case (Figure~\ref{fig:BoseHubbardPhaseDiagram}, trajectory 2). On the Mott side, we find two degenerate gapped (massive) modes. These are the particle-like and hole-like modes of the Mott insulator that are found using trial wave functions. The modes become gapless at the transition point, $r$=0. On the superfluid side, the antisymmetric combination of the particle and hole modes becomes the Goldstone boson, whereas the symmetric combination becomes the gapped (massive) Higgs boson.

Second, let us look at the generic case with no particle-hole symmetry (Figure~\ref{fig:BoseHubbardPhaseDiagram}, trajectory 1). We again see that there is a particle-like mode and a hole-like mode in the Mott insulator. However, this time at the transition only one of the modes becomes gapless, whereas the other mode remains gapped. On the superfluid side, the mode that became gapless at the critical point turns into the gapless sound mode.

Superfluid transitions that occur in $^4$He and in cold gases with no lattice can be described by the particle-hole asymmetric version of the $\phi^4$ model. As there is no mechanism to enforce particle-hole symmetry, we do not expect Lorentz invariance or a Higgs mode.  However, generically there should be a $K_2$ term in Equation~\eqref{action}, and thus we should still expect the existence of a gapped branch. However, there are two issues. First, as $K_2\rightarrow 0$ the gap at the transition becomes larger and larger $\omega_{k=0} \sim K_1/K_2$. Second, the gapped branch tends to be unstable due to the decay into sound modes and hence we should not expect any sharp signature. 

\subsection{Creation of the Higgs excitation}
We now consider the question of how to observe the Higgs mode in the particle-hole symmetric case.  This question has two parts: (1) Does the decay of the Higgs mode into the low-lying Goldstone modes render the Higgs mode unobservable; and (2) how does one couple to the Higgs mode? The observability of the Higgs mode, especially in 2D, was investigated by a number of authors~\cite{Chubakov1994, Sachdev1999, Zwerger2004,Lindner2010,Altman2002}.  The question was resolved in Reference~\onlinecite{Podolsky2011}, which argued that previous references considered different ways of driving the Higgs mode and hence obtained different answers. 

Consider two different representations of fluctuations of the order parameter $\Psi$ around its equilibrium broken value $\Psi_0$, in Eq.~\eqref{action}. In one representation, variations of the order parameter in longitudinal direction $u_L$ and transverse direction $u_T$ to the chosen directions, i.e., along and perpendicular to the vector $\Psi$ shown in Figure~\ref{Fig:mexhat}, are
considered to be the normal coordinates for the collective modes. In the second representation, radial coordinates $\eta$ and $\phi$ are used, which are the same as the $\delta_a$ and $(\Psi_0)^{-1}\delta_{ph}$ in Equation~\ref{eq:cartesian},
\begin{align}
\phi&=\text{Arg}(\Psi-\Psi_0) & \eta&=|\Psi|-\Psi_0.
\end{align}
In an expansion of the energy in terms of such variables, one finds that the leading term coupling $\eta$ and $\phi$ is $\eta \, |\nabla \phi |^2$. This is as it should be, as energy can depend only on the gradients of the phase variations to any order. But in the first manner of parameterization, the leading coupling term is $r ~ u_L ~u_T^2$. In calculations starting with such a parameterization, one finds that the correlation $\chi_{u_Lu_L}(\omega) = - \text{Im} \langle [u_L(\omega), u_L(-\omega)] \rangle \sim \omega^{-1}$ for long-wavelengths in $d=2$ and has a logarithmic divergence in $d=3$. On the other hand, using the radial parameterization, the Higgs readily appears as a peak in the correlation $\chi_{\eta \eta}(\omega) = - \text{Im} \langle [\eta(\omega), \eta(-\omega)]\rangle$. Unlike $\chi_{\delta_r \delta_r}(\omega)$, at low frequencies $\chi_{\eta \eta}(\omega) \sim \omega^{d+1}$, and hence the Higgs is not hidden by the low-frequency divergence~\cite{Podolsky2011}. This low-frequency behavior of $\chi_{\eta \eta}(\omega)$ may be obtained easily from the kinematic considerations of the decay, at long wavelengths, of the Higgs mode $\eta$ into a pair of Bogoliubov modes $\phi$, each with a dispersion $\omega \propto q$ through the coupling $\eta \, |\nabla \phi|^2$. 

In the experiments done so far, the correlation $\chi_{\eta \eta}(\omega)$ cannot be accessed directly, and an indirect connection with such a correlation through lattice modulation spectroscopy is used. The strategy for the spectroscopy consists of three steps: (1) Prepare a superfluid in an optical lattice. (2) Perform modulations of the depth of the optical lattice at a particular frequency. Because the hopping matrix element $J$ is exponentially sensitive to the lattice depth, this corresponds primarily to modulations of the KE operator. (3) Measure the total energy absorbed by the system~\cite{Endres2012} due to the initial modulation. As the measurement step is destructive, at the end of the three-step sequence, the atoms are thrown out of the trap. To obtain the next data point, the steps are repeated with newly cooled atoms and a different modulation frequency. Therefore, what is being measured experimentally is the total energy absorbed by the lattice gas in response to periodic modulations of the KE. In the case of weak drive, the total energy absorbed is related to the KE -- KE correlator via
\begin{align}
\Delta E (\omega)= \pi \delta^2 n_\text{mod} \text{Im} \left[ 
\int_0^\infty 
\frac{i}{\hbar} 
\langle [\text{KE}(t),\text{KE}(0)] \rangle 
e^{i \omega t} \, dt
\right],
\end{align}
where $\omega$ is the drive frequency, $\delta$ is the amplitude of the KE modulations, and $n_\text{mod}$ is the total number of modulations performed~\cite{Endres2013}. The idea is that for modulation frequency at the natural resonances of the superfluid state, the absorption peaks. 

Although lattice modulation spectroscopy couples to the Higgs mode, it does not do so cleanly, in the sense that KE modulations also excite particle-hole excitations. Therefore, the interpretation of the experiment relies on the comparison to analytical and numerical calculations with  Hamiltonian parameters that are known to a high accuracy. We now summarize the kinds of comparisons that can be made, before describing the experimental situation in the next subsection.

There are three choices of theoretical calculations that are available for comparisons. These are (1) mean-field and cluster mean-field calculations of the modulation dynamics in the Bose-Hubbard model~\cite{Endres2012}, (2) calculations of the $\chi_{\eta \eta}$ response in the continuum model of Eq.~\ref{action}~\cite{Podolsky2011, Podolsky2012}, and (3) the calculation of the $\chi_{\text{KE}\, \text{KE}}$ in the Bose-Hubbard model, using quantum Monte Carlo~\cite{Pollet2012,Gazit2013,Chen2014}. 

Choice (1) faithfully models the drive, the detection scheme and the trap that is present in the experiment, but it suffers from the fact that it does not calculate the decay of the Higgs mode into Goldstone modes accurately. It does a good job of capturing features seen in the experiment, but it is not applicable near the phase transition point. 

Choice (2) has the advantage of working with the universal model that captures the long wavelength physics near the phase transition point. However, it has the disadvantage that the relationship between what is being observed and what is being computed is not straightforward.

Choice (3) should have all the advantages of choice (1), but should also be able to capture the decay of the Higgs modes into the low-lying Goldstone modes. It is applicable both outside and inside the quantum critical regime. Specifically, numerical calculations obtain the imaginary frequency correlator $\langle \text{KE}(\tau),\ \text{KE}(0) \rangle- \langle \text{KE} \rangle^2$ using quantum Monte Carlo and then use an inverse Laplace transform to obtain the energy absorption line shape. Unfortunately, a direct comparison of Monte Carlo calculations with the experiment has yet to be made, as experimental details, such as the properties of the trap, have not been systematically taken into account in the calculations.

\begin{figure*}
\includegraphics[width=0.6\textwidth]{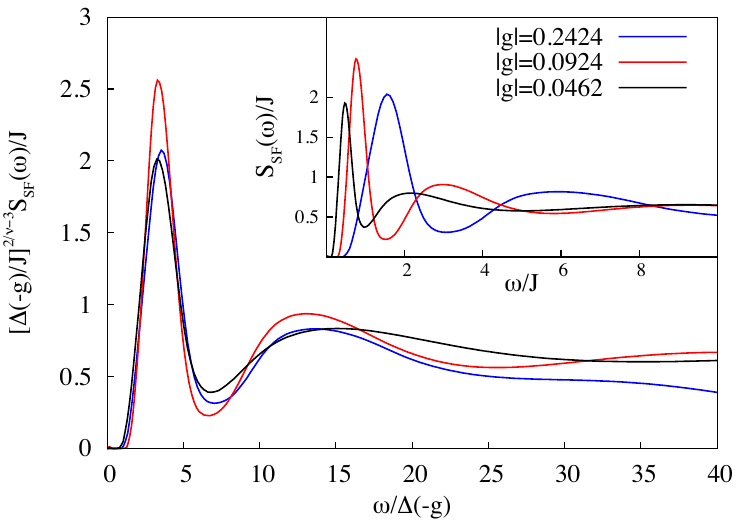}
\caption{
The spectral function $\text{Im} \, \chi_{\text{KE}\,\text{KE}}(\omega)=S_\text{SF}(\omega)$ as a function of the frequency $\omega$ showing the Higgs excitation in the 2D Bose-Hubbard model. The spectral function was obtained using quantum Monte Carlo calculations for several values of $g=(U-U_c)/J$ in the superfluid phase near the particle-hole symmetric quantum critical point. The inset shows the un-rescaled spectral functions while the main plot shows the collapse of the spectral functions upon the rescaling of the axes. Adapted from Reference~\cite{Chen2014}.
}
\label{fig:SpectralFunction}
\end{figure*}

Near the quantum critical point of the Mott-superfluid transition, analytical calculations of $\text{Im} \, \chi_{\eta\, \eta}(\omega)$~\cite{Podolsky2011, Podolsky2012} and numerical calculations of $\text{Im} \, \chi_{\text{KE}\, \text{KE}}(\omega)$~\cite{Pollet2012,Gazit2013,Chen2014} show that the Higgs line shape develops a characteristic broad maximum at Higgs mass frequency. Moreover, as one approaches the superfluid-Mott insulator quantum critical point from the superfluid side, the absorption line shape becomes universal. By universal, it is meant that the absorption line shapes obtained for various deviations from the critical point $\delta=(U-U_c)/J$ should be collapsed onto each other using a one-parameter scaling function $\Phi(\omega/\Delta)$
\begin{align}
A_\text{radial}(\omega) \sim C + A \Delta^{3-2/\nu} \Phi(\omega/\Delta),
\end{align}
where $\Delta$ is the gap scale $\Delta \sim |\delta|^\nu$, and $\nu=0.6723(3)$ is the correlation length critical exponent~\cite{Hasenbusch1999,Hasenbusch2001}. This universality in the line shape, obtained using a quantum Monte Carlo calculation of the KE linear response to KE modulations, is plotted in Figure~\ref{fig:SpectralFunction}, and is only reasonably good. 

\subsection{Experimental observation of the Higgs mode in ultra cold atoms}

The first experiments to study the excitations of bosonic superfluids on a lattice, with an attempt to observe the Higgs excitation, were reported in Refernces~\cite{Stoferle2004, Schori2004}. These experiments introduced the basic three-step strategy mentioned above and applied it to isotropic 3D lattices~\cite{Stoferle2004, Schori2004} and to 1D lattices and anisotropic lattices in the 1D-3D crossover regime~\cite{Stoferle2004}. The scheme for detecting the energy absorbed consisted of (1) ramping down the lattice to an intermediate depth to transfer the gas deep into the superfluid regime, (2) letting the gas thermalize, (3) turning off the lattice and trap potentials and letting the gas expand, and (4) taking an image of the interference pattern in the expanding cloud of atoms. The interference patterns observed in these images are related to the phase coherence across the atom clouds (following thermalization) and hence their temperatures, which is a measure of the energy absorbed. 

\begin{figure*}
\includegraphics[width=\textwidth]{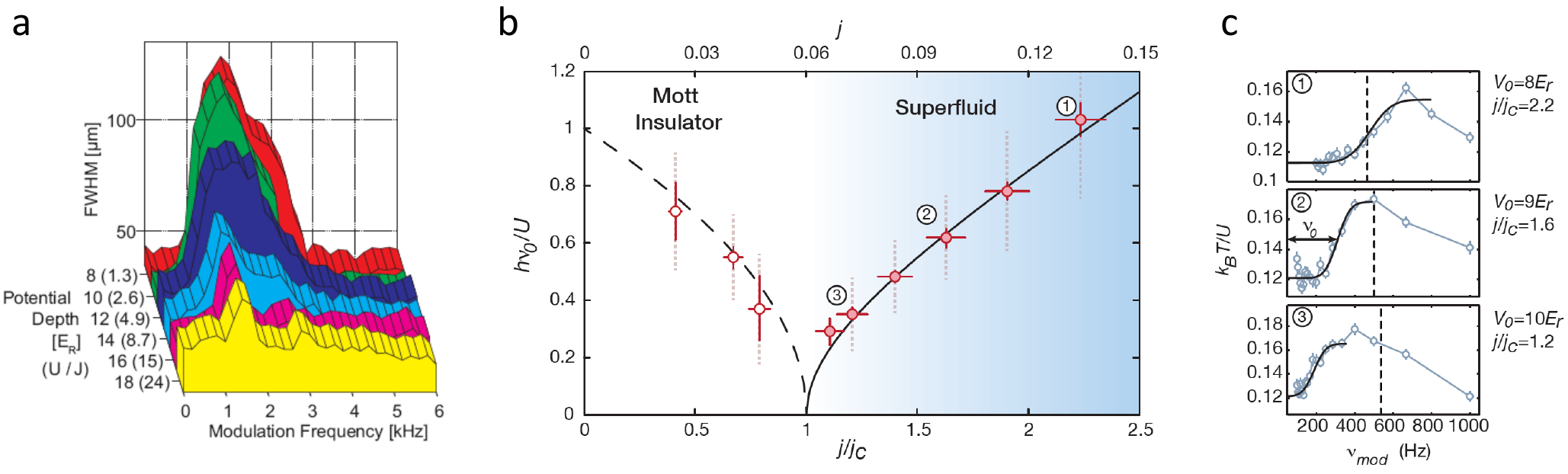}
\caption{
Experimental data from ultra cold atom groups showing the results of lattice modulation spectroscopy. {\bf a} Modulation spectra obtained in Reference~\cite{Stoferle2004}, showing the full width at half-maximum (FWHM) a measure of the energy absorbed by the atom gas, as a function of the lattice modulation frequency and the lattice depth. A pair of peaks can be seen at energy scale $\sim U$ and $\sim 2U$ (see text). {\bf b \& c} Lattice modulation data from Reference~\cite{Endres2012}. The three plots on the right show the temperature of the sample following lattice modulations, as a function of modulation frequency, at three different values of the lattice depth (as indicated). In each spectrum there appears a step-like absorption edge, associated with the Higgs mass (see text). The left hand plot shows the frequency associated with the absorption-edge feature as a function of the lattice depth (measured in $j/j_c$ where $j=J/U$ and $j_c=(J/U)_\text{critical}$). The points represent experimental data, the solid line shows the mass of the Higgs obtained from mean-field calculations and the dashed line shows the energy of the lowest particle-hole excitation. One can clearly see the softening of the Higgs mass on approach to the quantum critical point, thus allowing the identification of the absorption edge with the Higgs mass.
}
\label{fig:ExperimentalData}
\end{figure*}

To analyze the data, the full width at half maximum (FWHM) of the central peak in the interference images was measured and plotted as a function of the modulation frequency (see Fig.~\ref{fig:ExperimentalData}a for the 3D case). At large $U/J$, in the Mott-insulating phase, the FWHM shows a clear pair of peaks at $\omega \sim U$ and $\omega \sim 2U$. The lower peak can be ascribed to a single double occupancy-hole (particle-hole) excitation, while the higher peak corresponds to a pair of these excitation. As one approaches the Mott-insulator to superfluid transition, one would expect the lower peak to move to lower frequencies as the excitation gap of the Mott insulator vanishes. At the transition point, the peak should move all the way to zero frequency, before emerging again on the superfluid side as the Higgs excitation. Experimentally, the excitation spectrum indeed shows some changes as one approaches the Mott insulator - superfluid transition (for the 3D case the transition occurs at $U/J~16$~\cite{Capogrosso2007}). Unfortunately, the low frequency behavior is poorly resolved making it hard to see evidence for the closing of the excitations gap and the emergence of the Higgs gap. There is always structure at higher frequencies associated with $\omega \sim U$ scale. This structure may be related to the local physics of double occupancy-hole excitations.

Two subsequent experiments have since been done. In one, spatially nonuniform lattice modulations were used to to impart a momentum of $\frac{\pi}{\sqrt{2} a}(1,1,0)$ to the quasi-particle modes that were being excited~\cite{Bissbort2011}. The measurement procedure was similar to that of the earlier experiments: the optical lattice was ramped down and then the cloud was allowed to expand in order to produce a time of flight image. In order to interpret their results, sophisticated numerical modeling of the experiments was done. The results of the modeling were consistent with experimental data. As the experimental probe measured the response only at high momenta, near the Brillouin zone boundary, the results obtained were not directly related to the physics of the long-wavelength Higgs mode.  

The other experiment repeated the Esslinger experiments on 2D lattices using uniform lattice modulations, but it achieved much higher sensitivity in the measurement step by using a ``quantum gas microscope" to image the atoms in the optical lattice with single-site resolution~\cite{Bakr2010,Sherson2010,Endres2012}.
Specifically, to image the atoms, the optical lattice is slowly ramped up to confine the atoms to individual lattice sites. Once the atoms are confined, they are imaged using florescence imaging through a microscope with a large numerical aperture. The resulting image shows the parity of the number of atoms in each lattice site, i.e., bright spots for sites with odd numbers of atoms and dark spots for sites with even numbers of atoms (the imaging procedure was unable to measure the atom number because of a photo-assisted recombination process that kicked out pairs of atoms from the lattice, and thus emptied the sites with an even number of atoms, during the imaging procedure). To obtain the energy absorbed by the gas during lattice modulations, the sample temperature was reconstructed from the radial distribution of the even and the odd sites. 

The employment of the quantum gas microscope resulted in a qualitative improvement in the data. Modulation spectra obtained by this method showed a sharply rising feature at low frequencies indicative of a gap (see Figure~\ref{fig:ExperimentalData}c). The size of this gap depends on the proximity to the quantum critical point. As $J/U$ decreases, this gap shrinks, reaching its smallest value at the quantum critical point, before expanding again on the Mott insulating side (see Figure~\ref{fig:ExperimentalData}b). Theoretical modeling  shows that both the value of the gap and its dependence on $J/U$ matched the mass of the Higgs excitation on the superfluid side and the particle-hole gap on the Mott insulating side, thus identifying the origin of this gap. 

\begin{figure*}
\includegraphics[width=\textwidth]{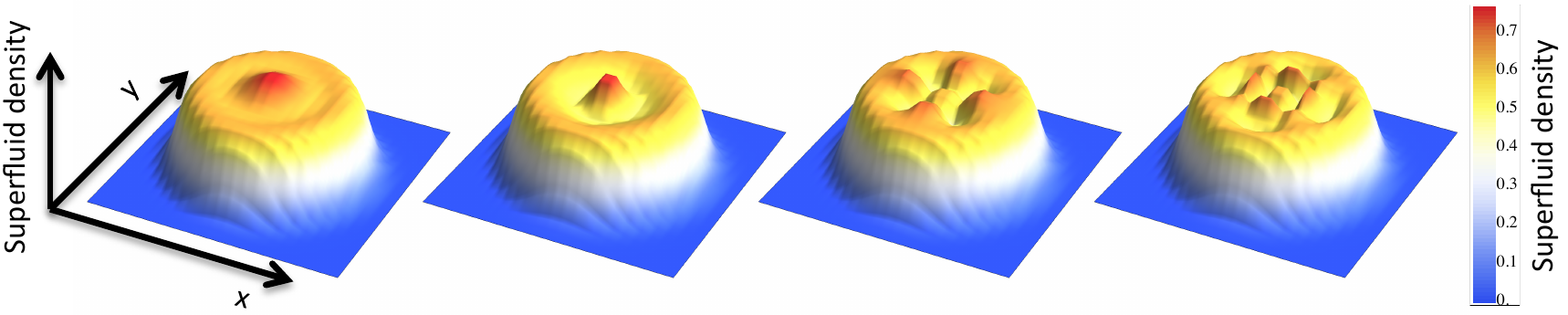}
\caption{
Higgs drum modes in a trap, taken from data in Reference~\cite{Endres2012}. The four plots show the superfluid density as a function of position in the trap, after lattice modulations at frequencies corresponding to the lowest four Higgs modes. The plots were obtained by numerically evolving the product wave function of Equation~\ref{eq:PsiTrial}.  
}
\label{fig:HiggsDrum}
\end{figure*}

The question arises, why did the experiments observe a threshold feature as opposed to a peak at the Higgs mass scale? There are two possible reasons. One is that the experiment also observes absorption by the continuum of particle-hole excitations that have a similar threshold as the Higgs energy. The other reason is the nonuniform boson density due to the trapping field. In the center of the trap one finds the density to be one boson per site, i.e.,  particle-hole symmetry. As one moves toward the edges of the trap, the density of bosons decreases, and the massive mode gets more massive and is also damped. The absorption by these massive modes at the edges of the trap probably stretches the Higgs peak into a line-shape with a threshold. 
This local density description can be refined by constructing the normal modes in a trap. The Higgs excitations appear in the form of trap modes that appear similar to the modes of a drum head (see Figure~\ref{fig:HiggsDrum}). The lowest trap mode occurs slightly above the Higgs mass in free space. It is followed by a discreet spectrum of higher frequency modes. Uniform lattice modulations couple to all trap modes with zero angular momentum. Thus, during lattice modulation spectroscopy one observes the absorption into a number of trap modes in contrast to the absorption into the single $k=0$ mode in free space. Monte Carlo analysis~\cite{Pollet2012} and experimental data~\cite{Endres2012}, suggest that the Higgs modes are too short-lived for observation of  individual trap frequencies. The broadened spectrum of trap modes resembles the observations.

In summary, with the refinements that were introduced in Ref.~\cite{Endres2012}, ultracold atom experiments present evidence for the existence of the Higgs mode in lattice superfluids. As the lattice modulation spectroscopy does not cleanly couple to the Higgs mode, the evidence is based on a combination of experimental measurements and comparison to analytical and numerical calculations. 

\subsection{Outlook}
In the future we may expect experiments on ultracold atoms to provide us with a wealth of new data to help understand many-body phenomena. A possible way to observe the Higgs mode directly would be to change the driving mechanism. Instead of coupling to the Higgs mode via the KE, one could couple to it via long wavelength density or current modulations. One may then generate the Higgs directly through the process depicted in Figure~\ref{coupling}c, although this process is implicated in the experiments that have already been done. The direct exceptions of Higgs may be accomplished by gradients of the chemical potential across the lattice or by generating a synthetic gauge field~\cite{Lin2009} to couple to the current. This type of coupling, in addition to being cleaner, could be used to extract the dispersion relation of the Higgs mode. These techniques may require shallower lattice potentials, which create further problems in controlling the experiment.

Although current state-of-the-art experiments can approach the quantum fluctuation regime in 2D lattice bosons, we expect that ultracold atoms will provide quantitatively reliable data on the 1D, 2D, and 3D Mott insulator superfluid quantum critical points, with improved measurement precision and lower temperatures.  Such experiments may answer the questions about universal features of the Higgs excitation near criticality.  

Another area that could be addressed by ultracold atom experiments is the fate of the Higgs excitation in fermionic superfluids. In the weakly interacting fermionic superfluids, the Higgs mass coincides with the edge of the particle-hole excitation spectrum, making the Higgs hard to observe. However, interactions  bring down the Higgs mass in relation to the particle-hole continuum. This raises the question of whether the Higgs can be observed in experiments with strong interactions (i.e., near a Feshbach resonance), or dipolar interactions (i.e., in polar atomic or molecular gases). Alternatively, strong confinement can be used to protect the Higgs mode from decay, as pointed out in Reference~\onlinecite{Bruun2014}.

The development of artificial static and dynamic gauge fields~\cite{Banerjee2012} also lies in the future and holds promise for further probing of the amplitude and phase modes. 

\section{Higgs excitation in magnetic materials}
We have so far described the appearance of the Higgs excitation in superfluid and superconducting systems in which the condensate is made up of particles or pairs of particles.  Magnetic systems offer an alternative, and in some sense richer, setting in which the condensate consists of the magnetization of the system. At the transition to the ordered phase, the magnetization spontaneously breaks rotation symmetry as it acquires a specific direction in space. 

In a model of interacting spins that can be mapped to the xy model, the Goldstone excitations are the spin-waves that correspond to small fluctuations in the direction of the magnetization. The Higgs excitations are the spin-waves in which the magnitude of the magnetization fluctuates. The existence of both kinds of spin-waves has been theoretically known and studied for a long time by inelastic polarized neutron scattering and inelastic light scattering. However, recently, the identification of the collective modes as Goldstone-like and Higgs-like has been made by studying a system near its quantum critical point: The idea is to see that on the magnetically ordered side of the critical point, one should observe both massless and massive spin-waves, with the mass being a function of the distance to the critical point. 

\begin{figure}
\includegraphics[width=8cm]{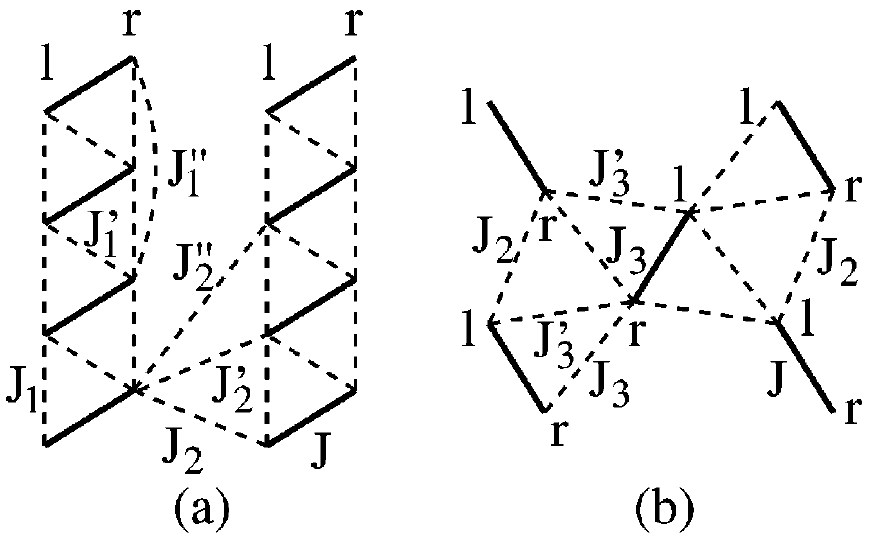}
\caption{The Heisenberg like model of \(\text{TlCuCl}_3\). The vertices represent the magnetic moments of the Copper atoms, the thick solid lines represent the position of the dimers as well as the intra-dimer coupling $J$, the thin solid lines represent the inter-dimer couplings $J_1$, $J_2$, etc. Adapted from Reference~\cite{Matsumoto2004}.}
\label{fig:MatsumotoModel}
\end{figure}

Around 2003, it was found that applying hydrostatic pressure to the compound \(\text{TlCuCl}_3\) induces a quantum phase transition between a nonmagnetic state and an antiferromagnetic state, at the critical pressure $P_c=0.42\,\text{mbar}$~\cite{Oosawa2003, Tanaka2003, Ruegg2004, Oosawa2004, Goto2004}. Structurally, \(\text{TlCuCl}_3\) consists of one-dimensional chains of dimers, with weak interchain couplings (see Figure~\ref{fig:MatsumotoModel}). The quantum phase transition is driven by the competition between the intradimer couplings (which favor the nonmagnetic dimer singlet ground state) and the interdimer couplings (which favor the antiferromagnetic ground state). Applying hydrostatic pressure drives the phase transition by tuning of the relative strength of the inter- and intradimer coupling. 

\begin{figure}
\includegraphics[width=8cm]{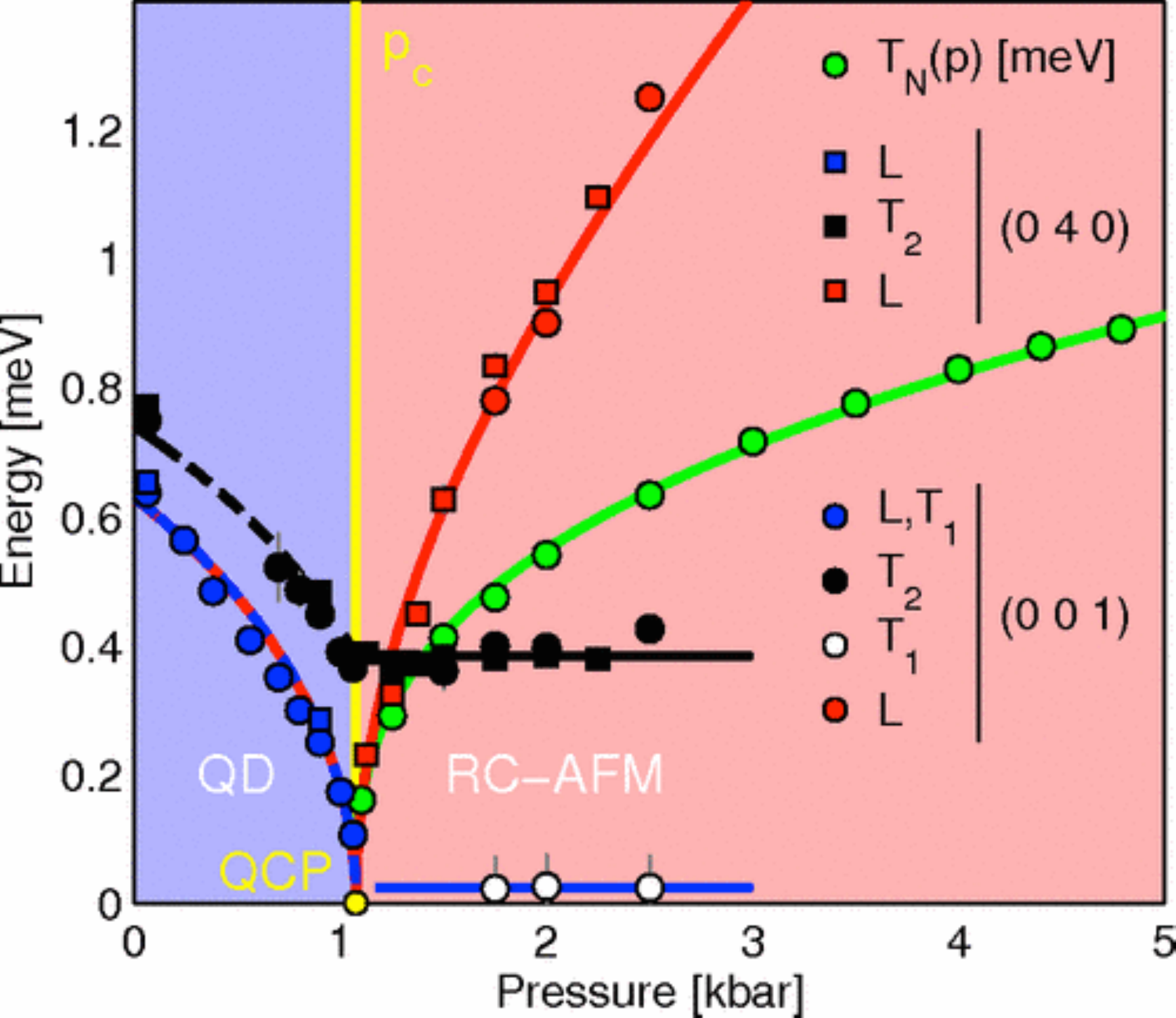}
\caption{The mass of the three spin-wave modes of $\text{TlCuCl}_3$ obtained from inelastic neutron scattering. The figure shows Neel temperature (green points) and the mass and the three spin-wave modes (the remaining points) as a function of the hydrostatic pressure (see text). The solid lines represent theoretical fits from the minimal model of Equation~\ref{eq:RueggModel}. Abbreviations: QCP, quantum critical point; QD, quantum disordered phase; RC-AFM, renormalized classical antiferromagnetic. Adapted from Reference~\cite{Ruegg2008}.}
\label{fig:RueggSpectra}
\end{figure}

A detailed study of the spin-wave modes of \(\text{TlCuCl}_3\) in the vicinity of the quantum critical point was performed by R{\"u}egg et al.~\cite{Ruegg2008} using inelastic neutron scattering. The neutron-scattering data (see Figure~\ref{fig:RueggSpectra}) show three massive spin-wave modes in the nonmagnetic phase. The mass of all three modes decreases as the quantum critical point is approached. At criticality, two of the modes become massless, whereas the third mode remains massive. As one pushes into the antiferromagnetic phase, 
the massive mode remains massive, one of the massless modes remains massless (the Goldstone spin-wave), and the other massless mode acquires a mass (the Higgs spin-wave). 

In order to make a conclusive identification of the various spin-wave modes, R{\"u}egg et al. used a Heisenberg-like model of magnetism that was introduced in Reference~\onlinecite{Matsumoto2004}. The original model consists of intradimer couplings $J$ and interdimer couplings $J_1$, $J_2$, etc., as shown in Figure~\ref{fig:MatsumotoModel}. Here we present a minimal version of this model, which is sufficient to analyze the neutron-scattering data in Ref.~\cite{Ruegg2008}:
\begin{align}
H=J\sum _a S_{l,a}\cdot S_{r,a}-J_{xx}\sum _a S_{l,a}^xS_{r,a}^x+J_2\sum _{\langle a,b\rangle }S_{l,a}\cdot S_{r,b},
\label{eq:RueggModel}
\end{align}
where $l$ and $r$ are the left- and right-hand sites of the dimer, $a$ and $b$ are the dimer positions, $\langle a,b\rangle$ indicate the nearest dimers, and $S$ is a spin 1/2 operator. The $J$ interaction couples the two sites of the same dimer and therefore when it is the dominant interaction each dimer goes into a singlet ground state and the system becomes nonmagnetic. However, when the intradimer coupling \(J_2\) becomes comparable to the \(J\) coupling, the system develops 3D antiferromagnetic order. Importantly, the weak $J_{xx}$ coupling models the anisotropy of the dimer bond. In \(\text{TlCuCl}_3\), the  $J_{xx}$ term favors $y$ and $z$ triplets on dimers, and hence introduces an ``easy plane'' (i.e. a ``hard axis'') anisotropy. The easy plane anisotropy changes the symmetry being broken from $\text{SU}(2)$ to $\text{U}(1)$. The 3D character arises due to the inter-chain couplings that are not captured by the minimal model. Analysis of the minimal model at the mean-field level is, however, sufficient to capture the main experimental observations. 

To get a picture of the phase diagram, as well as that of the excitations supported by the minimal model, consider the following product wave function 
\begin{align}
|\psi \rangle =\prod _{a}\left(\sqrt{1-|\alpha_a|^2-|\beta_a|^2-|\gamma_a|^2} \, |s\rangle_a
+\alpha_a |t_1\rangle_a + \beta_a |t_2 \rangle_a + \gamma_a |t_3\rangle_a \right).
\end{align}
Here, \(|s\rangle_a =\frac{1}{\sqrt{2}}|\uparrow \downarrow - |\downarrow \uparrow \rangle_a\)
represents a singlet on the dimer $a$, and $|t_1\rangle_a =\frac{1}{\sqrt{2}}|\uparrow
\downarrow + \downarrow \uparrow \rangle_a$, $|t_2\rangle_a =\frac{i}{\sqrt{2}}|\uparrow \uparrow
+ \downarrow \downarrow \rangle_a$, and $|t_3\rangle_a =\frac{1}{\sqrt{2}}|\downarrow \downarrow-\uparrow \uparrow
 \rangle_a$ are the three triplets. We choose the triplets so that $|t_1\rangle_a$
and $|t_2\rangle_a$ have the same energy, and $|t_3\rangle_a$ lies just above in energy due to the
easy-plane anisotropy. The parameters $\alpha_a$ and $\beta_a$ play the role of the real and imaginary part of the superfluid order parameter $\phi$. Hence the nonmagnetic ground state corresponds to $\alpha_a=\beta_a=\gamma_a=0$ and the magnetic ground state corresponds to $|\alpha_a|^2+|\beta_a|^2 \neq 0$.

The spin-wave theory, at the mean-field level, can be obtained by extremizing the matrix element $\langle \psi | i \partial_t-H | \psi \rangle$ with respect to $\alpha_a$, $\beta_a$, and $\gamma_a$, which results in a set of equations that are analogous to the usual Bogoliubov
equations. The resulting zero momentum modes were plotted as solid lines in Fig.~\ref{fig:RueggSpectra}. In the nonmagnetic state, there are three massive modes. Two modes with identical energy corresponding to the triplets $|t_1\rangle$ and $|t_2\rangle$; and another mode with slightly higher energy corresponding to the triplet $|t_3\rangle$. As we approach the quantum critical point, the gap for the two degenerate modes goes to zero, whereas the third mode remains gapped. On the magnetic side, the two degenerate modes split into a gapless Goldstone mode and a gapped Higgs mode, whereas the third mode remains gapped. Hence, using the simple model, the modes observed in neutron scattering could be conclusively identified. We note in passing, that in the absence of an anisotropy, on the nonmagnetic side the third mode would become degenerate with the other two modes, whereas on the magnetic side it would become another Goldstone mode.

{\it Disclosure Statement}: The authors are not aware of any affiliations, memberships, funding, or financial holdings that might be perceived as affecting the objectivity of this review.

{\it Acknowledgements}: We wish to acknowledge useful discussions with and suggestions from Daniel Arovas, Assa Auerbach, Andrew Cohen, Manuel Endres, Thierry Giamarchi, Alan Goldman, Marie-Aude M$\grave{e}$asson, Andrea Morales, Daniel Podolsky, and Lode Pollet. 

\bibliography{HiggsBib-v2-1}
\end{document}